\documentclass[a4paper,11pt]{article}
\usepackage{jcappub} % for details on the use of the package, please
\usepackage{latexsym}
\usepackage{graphicx}
\usepackage{psfrag}

\usepackage{mathptmx}       % selects Times Roman as basic font
\usepackage{helvet}         % selects Helvetica as sans-serif font
\usepackage{courier}        % selects Courier as typewriter font
\usepackage{type1cm}        % activate if the above 3 fonts are
\usepackage{makeidx}         % allows index generation
\usepackage{multicol}        % used for the two-column index
\usepackage[bottom]{footmisc}% places footnotes at page bottom

\def\apjl{Astrophys. J. Lett.}
\def\mnras{Mon. Not. Roy. Astron. Soc.}

\def\physrep{Phys. Rept.}
\def\apj{Astrophys. J.}
\def\apjs{Astrophys. J. Suppl.}
\def\aap{Astron. Astrophys.}
\def\araa{Annu. Rev. Astron. and Astrophys.}
\def\nar{New Atron. Rev.}
\def\prd{Phys. Rev. D}
\def\apjl{Astrophys. J. Lett.}
\def\mnras{Mon. Not. Roy. Astron. Soc.}

\def\physrep{Phys. Rept.}
\def\aap{Astron. Astrophys.}
\def\araa{Annu. Rev. Astron. and Astrophys.}
\def\nar{New Atron. Rev.}
\def\prd{Phys. Rev. D}

\title{Dark energy equation of state parameter and its evolution at low redshift}

\author[a,b]{Ashutosh Tripathi}
\author[a]{Archana Sangwan}
\author[a,1]{H. K. Jassal \note{Corresponding author.}}

\affiliation[a]{Indian Institute of Science Education and Research
Mohali,\\ SAS Nagar, Mohali 140306, Punjab, India.}
\affiliation[b]{Center for Field Theory and Particle Physics and 
Department of Physics,\\ Fudan University, 200433 Shanghai, China.}

\emailAdd{ashutosh\_tripathi@fudan.edu.cn}
\emailAdd{archanakumari@iisermohali.ac.in}
\emailAdd{hkjassal@iisermohali.ac.in}

\abstract {In this paper, we constrain dark energy models using a compendium of
observations at low redshifts. 
We consider the dark energy as a barotropic fluid, 
with the equation of state a constant as well the case where dark
energy equation of state is a function of time.  
The observations considered here are Supernova Type Ia data,
Baryon Acoustic Oscillation data and Hubble parameter
measurements.
We compare constraints obtained from these data and also do a combined
analysis.
The combined observational constraints put strong limits on variation of
dark energy energy density with redshift.
For varying dark energy models, the range of parameters preferred by
the supernova type Ia data  is in tension with the other low redshift distance
measurements.}

\begin{document}
\maketitle

\section{Introduction}
\label{sec::intro}

The acceleration of the cosmic expansion is one of
the most important discoveries in present day cosmology \cite{sn1,sn2,sn3,sn4,sn5,sn6,sn7,sn8,sn9}.
This acceleration requires that nearly three-quarters of the energy of
the universe is in  a component with a negative pressure, namely
dark energy.
A large number of models have been proposed in an attempt to explain
dark energy.    
The simplest of these models is the cosmological constant models and
this model is consistent with observations.
However, due to the fine tuning problem \cite{cc1,cc2,cc3,cc4,cc5,cc6,cc7,cc8,cc9,cc10,cc11,cc12}
models of dark energy have been proposed;  based on fluids and on
canonical and noncanonical scalar fields \cite{sf1,sf2,sf3,sf4}.
The condition for late time acceleration is that the equation of
state parameter of dark energy is $w<-1/3$, where w is the ratio of
pressure $p$ and the energy density $\rho$.
The equation of state parameter can be a constant or it can be
a function of time. 

A simple model of dark energy with a constant or a varying equation of
state parameter is that of a barotropic fluid.    
A parameterized form of w(z) is assumed for varying dark energy.   
The parameterization may, for example, be a Taylor series expansion
in the redshift, 
a Taylor series expansion in the scale factor or any other general
parameterization for $w(z)$ \cite{prm1,prm2,prm3,prm16,prm4,prm5,prm6,prm7,prm8,prm9,prm10,prm11,prm12,prm13,prm14,prm15}.
The parameters can then be constrained using different observations. 
For  detailed reviews, see \cite{rev1,rev2,rev3}.

In this paper we study dark energy modeled by a barotropic fluid, in
the context of present observations. 
We consider three different scenarios, namely, a constant equation of
state parameter of dark energy, and three different parameterizations of
dark energy  parameter with  a variable $w$.  
In this paper, we use type Ia supernova (SNIa) data 
\cite{sn1,sn2,sn3,sn4,sn5,sn6,sn7,sn8,sn9}, Baryon Acoustic Oscillation(BAO) 
data \cite{bao00,bao1, bao2, bao3,bao4,new_bao1, new_bao2} and Hubble parameter (H(z)) 
data \cite{hz1,hz2,hz3,hz4,hz5,hz6,hz7,hz8,hz9,hz10}. 
Other potential observational constraints include those from Gamma Ray Bursts
(GRB) data \cite{grb1,grb2,grb3,grb4,grb5},  angular size versus redshift data \cite{da1,da2}, Cosmic 
Microwave Background (CMB) data \cite{planck1,planck2}. 
The GRB and angular size versus redshift datasets, 
as yet, are not as restrictive as SNIa, BAO, H(z) or CMB \cite{planck1,planck2}.
In this paper, we concentrate on dark energy equation of state
parameter value and its evolution at low redshifts.
Since we are investigating the low redshift behavior of dark energy, 
we consider the SNIa, BAO and H(z) data for this analysis.
In particular, we investigate if one functional form of the dark
energy parameters is preferred over the others by observations.
We demonstrate that 
combined analysis from these datasets constrain the
allowed range of parameters significantly.
The constraints obtained from 
observations are consistent with a cosmological constant.

The paper is structured as follows.  
After introductory Section \ref{sec::intro}, in section
\ref{sec::bgeqns}, we review background cosmology and   
in section \ref{sec::obsv}, different observations are 
discussed and used to constrain different models.
Section \ref{sec::results}  presents our results and the last
section \ref{sec::conclusions} summarizes the  main results and
concludes.

\section{Cosmological Equations}
\label{sec::bgeqns}

A homogeneous and isotropic universe is  described by 
Friedmann equations which for a spatially flat geometry are given by 
\begin{eqnarray} \label{friedmann}
\frac{\dot{a}^2}{a^2} &=& \frac{8\pi G\rho}{3} \\ \nonumber
\frac{2\ddot{a}}{a}+\frac{\dot{a}^2}{a^2} &=& -{8\pi G P}
\end{eqnarray}
where $\rho$ is the total energy density and $P$ is the pressure of 
the universe\cite{geom}.
We have assumed speed of light $c$ to be
unity.
The total energy density is given by
\begin{eqnarray}
\rho &=& \rho_{m} + \rho_R + \rho_{DE}
%&=& \rho_c\begin{pmatrix}\Omega_r (\frac{a_0}{a})^4 +\Omega_n (\frac{a_0}{a})^3 + \Omega_v (\frac{a_0}{a})^3(1+w)\end{pmatrix}
\label{rho}
\end{eqnarray}
where the subscripts $m$, $R$ and $DE$ denote the non-relativistic,
relativistic (including radiation) and the dark energy components respectively.
The non-relativistic matter includes baryonic matter and dark matter.
The total energy density can also be expressed in terms of density
parameters  $\Omega_i$ = $\rho_i/\rho_c$, where \textit{i} indicates
the values  of different components of $\rho$ given in Eq.~\ref{rho}
and $\rho_c$  is the critical energy density of the universe, which is
given by  $\rho_c=3{H_0}^2/8\pi G$.
Since the relativistic component is subdominant at late times, 
for a spatially flat universe $\Omega_{m} + \Omega_{DE} \approx 1$.
In other words, the energy density of the universe comprises primarily of
the nonrelativistic matter and dark energy components.
Therefore, the first equation in Eq. \ref{friedmann} can be rewritten as
\begin{equation}
H^2 = H_0^2\left[\Omega_{m}(1+z)^3 +\Omega_{DE} \textrm{exp}\left\{3\int_0^{z} \frac{dz}{1+z}\left[1+w(z)\right]\right\}\right]  \label{hsquare}
\end{equation}
where $w$ is the equation of state parameter for dark energy component
and $z$ is the redshift.  

\begin{figure}[tbp]
\centering
\includegraphics[width=0.2\textwidth]{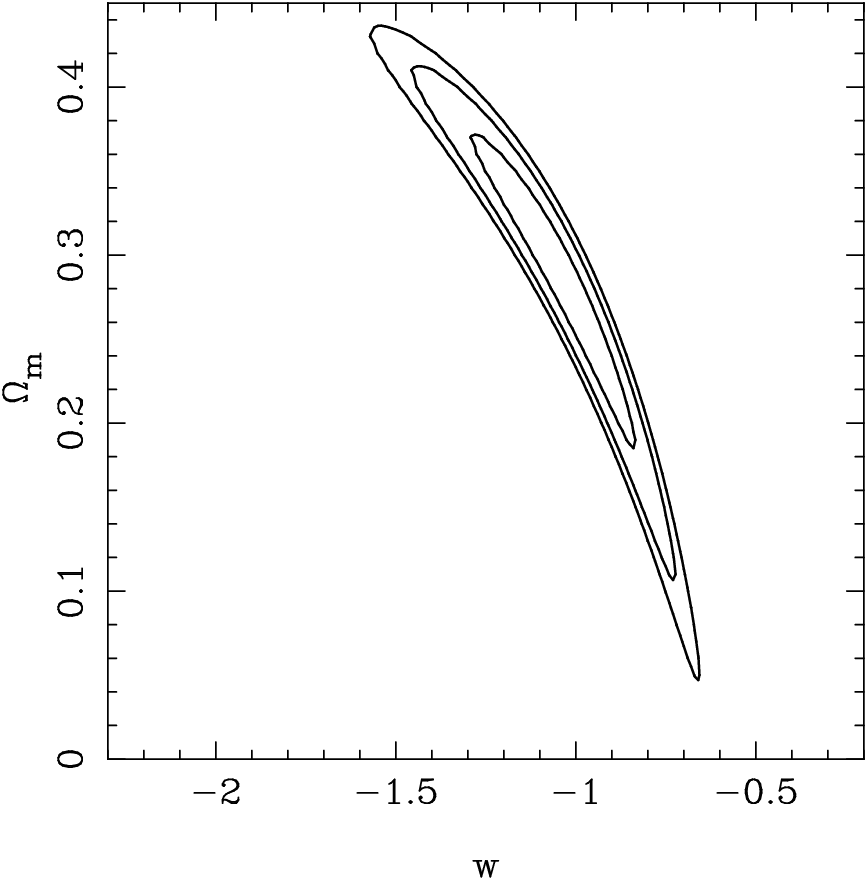} \includegraphics[width=0.2\textwidth]{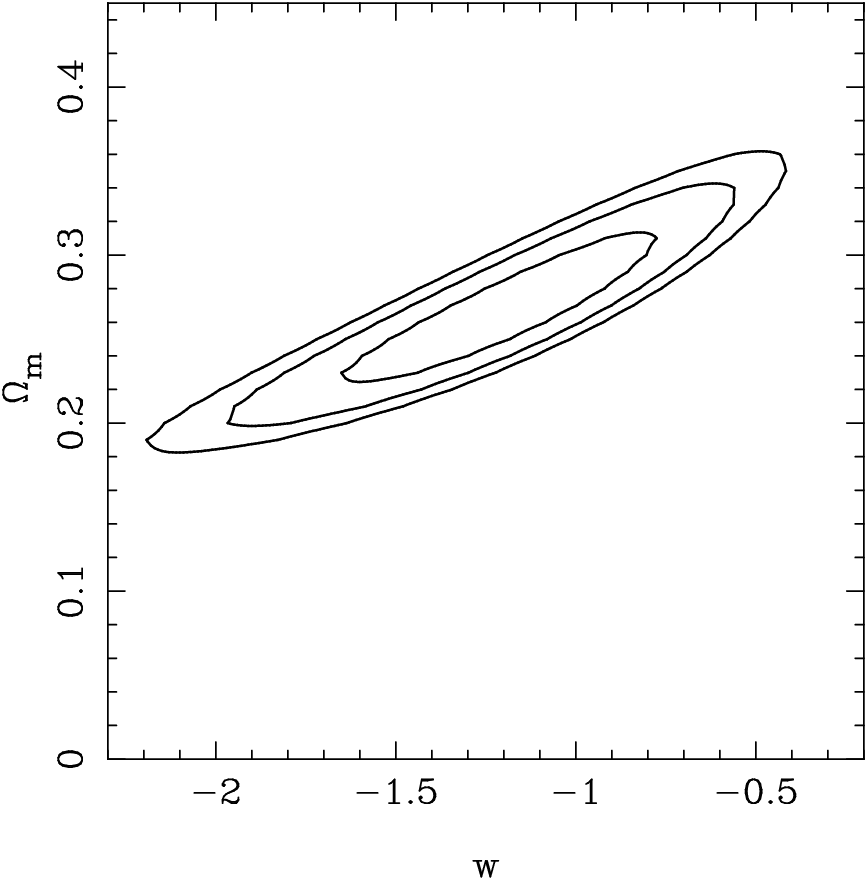}\includegraphics[width=0.2\textwidth]{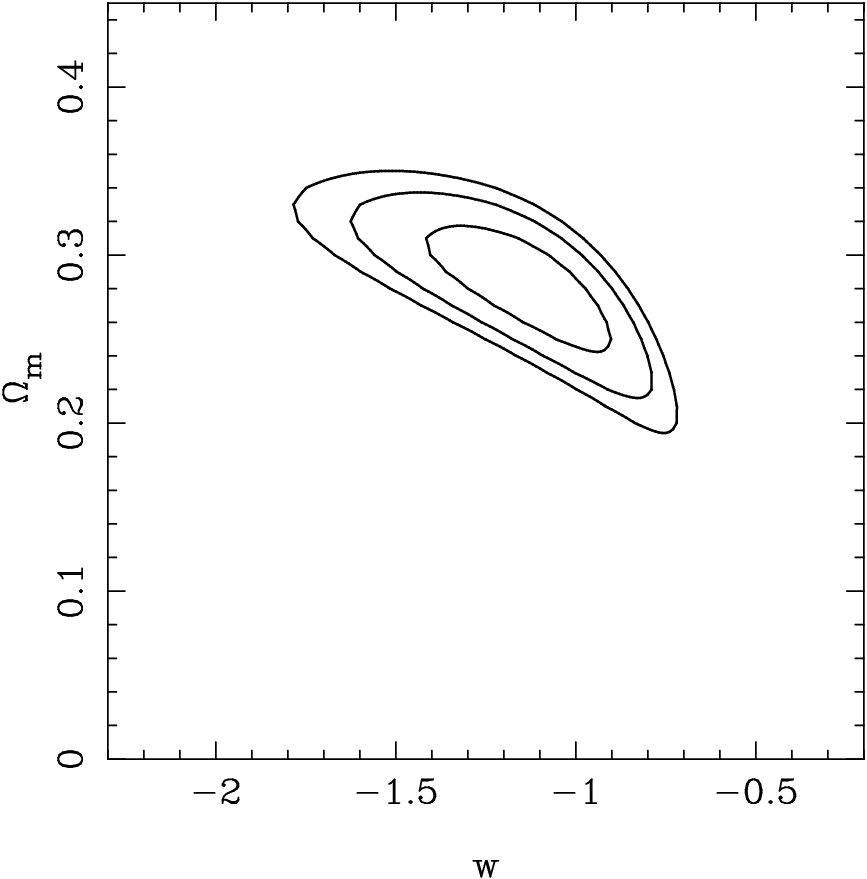} \includegraphics[width=0.2\textwidth]{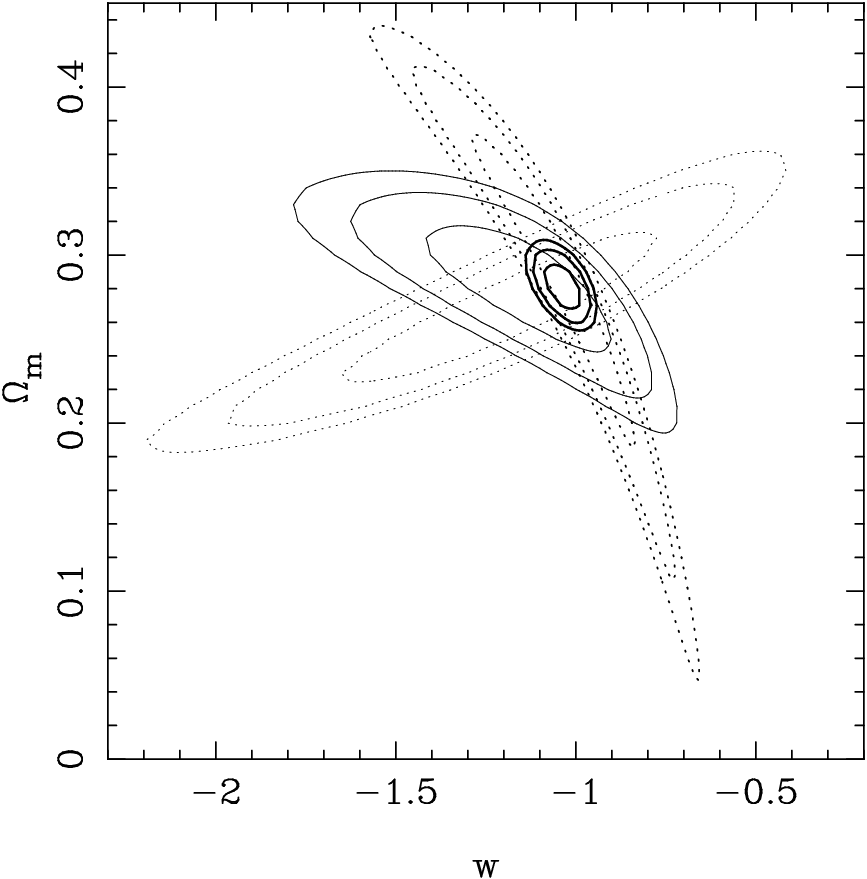}\\ 
\includegraphics[width=0.2\textwidth]{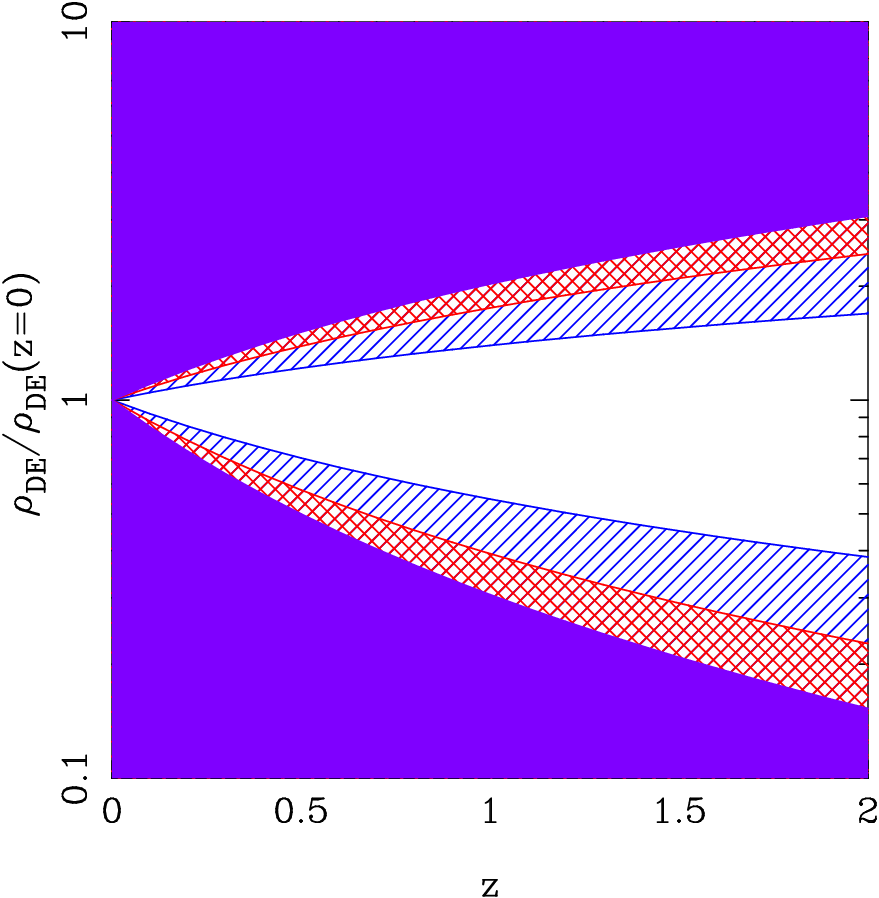} \includegraphics[width=0.2\textwidth]{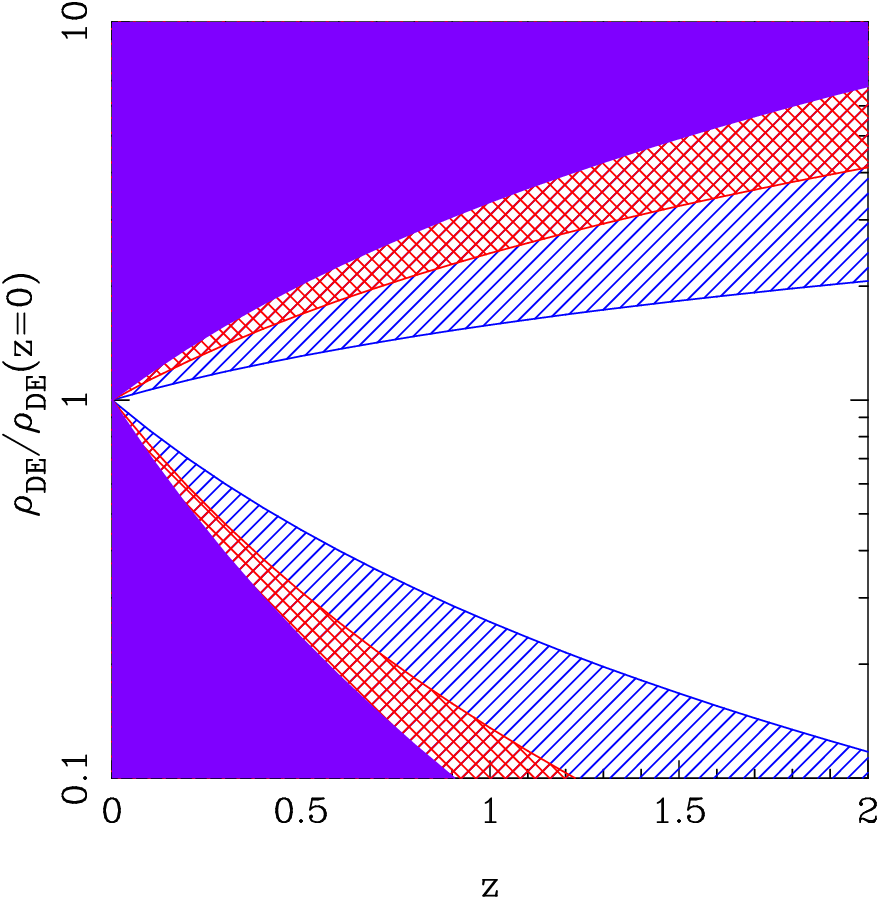} \includegraphics[width=0.2\textwidth]{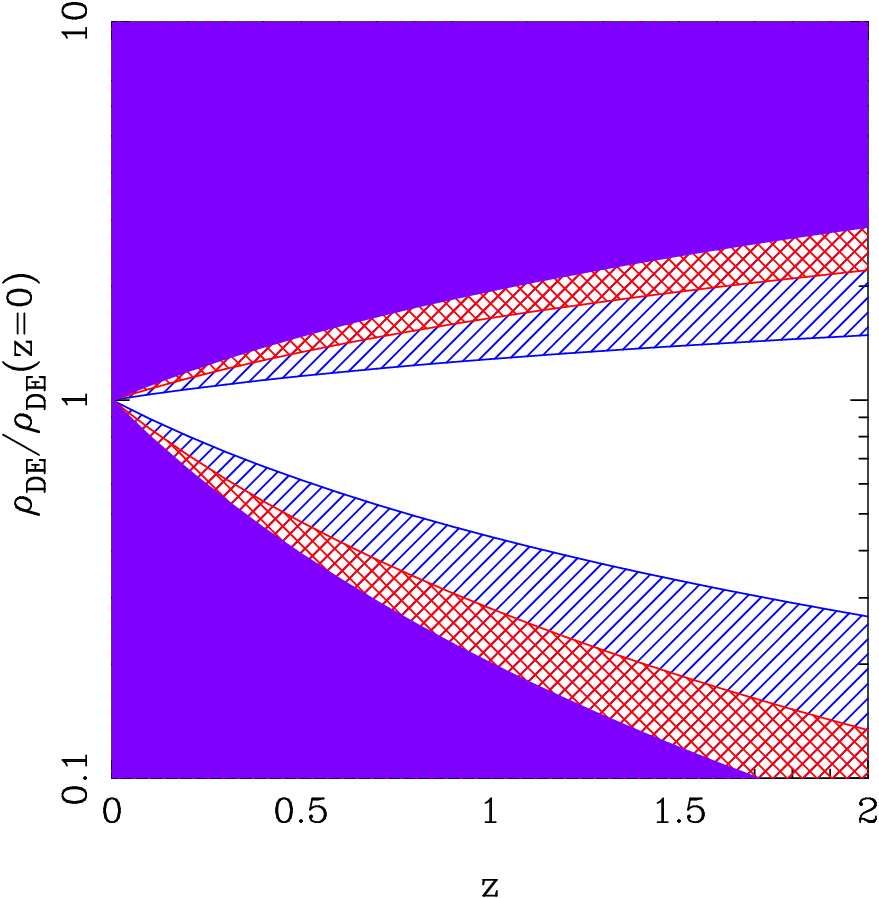} \includegraphics[width=0.2\textwidth]{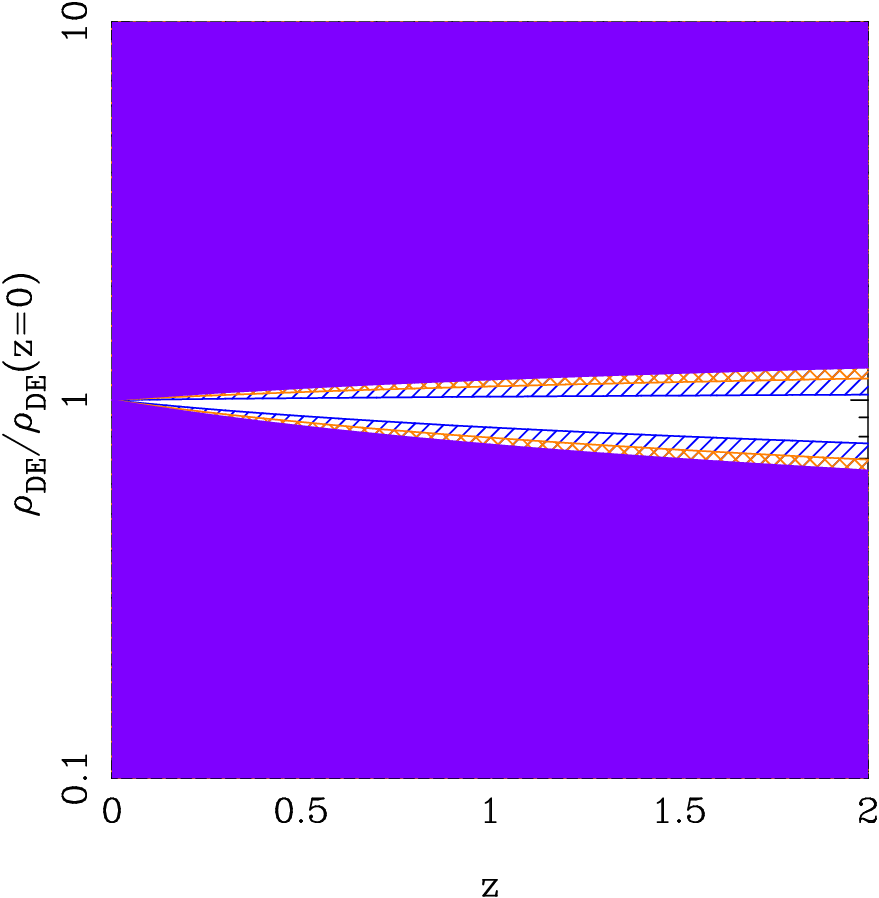}
%\end{tabular}
\caption{\label{fig::constw} The first row of the figure represents confidence  contours
  at $1 \sigma$, $2\sigma$ and $3\sigma$ in $\Omega_{m}-w$ plane. Going from 
  left to right, the results are for SNIa , BAO and H(z) data 
  respectively. The fourth plot represents the combined result. 
  The plots in the second row represents dark energy density versus redshift
  for the aforementioned data sets and their combination in the same
  order as before. The white region in the middle is the allowed range
  of dark energy density at 1$\sigma$ level and the regions with slanted lines
  (Blue) and hatched lines (red) allowed 2$\sigma$ and 3$\sigma$ ranges
  respectively. The solid blue region is ruled out.}  
\end{figure}

We consider two different models; one with a constant $w$ and another with a varying $w$. 
The varying equation of state parameter can be approximated by a
function of redshift.
Although many parameterizations have been proposed, we use the following  parametrizations in
this analysis: 
\begin{equation}
w(z) = w_0 + w'(z=0)\frac{z}{(1+z)^p}; \hspace{1cm} p = 1, 2
\end{equation}
and the logarithmic paramterization, given by
\begin{equation}
  w(z)=w_0 + w'(z=0) log(1+z)
  \end{equation}
where $w_0$ is the value of equation of state parameter at present 
and $w'(z=0)$ is the first order derivative of $w(z)$ at $z=0$ .
The redshift behavior is  different in  these two parameterizations. 
If $p = 1$ \cite{prm1,prm2}, the asymptotic value of w(z) at high redshifts is
$w(z=\infty) = w_0 + w'$ and for the parameterization $p = 2$,
$w(z=\infty) = w_0$ \cite{prm3}.
The equation of state parameter increases monotonically for the
logarithmic parameterization \cite{prm16}. 
The present value of the equation of state parameter is $w(0)=w_0$ for
all these parameterizations.

The evolution of dark energy density $(\rho_{_{DE}})$ with the expansion 
of the universe for these equation of state parameters are  then given by
\begin{equation}
\frac{\rho_{_{DE}}}{\rho_{_{DE_0}}} = (1+z)^{3(1+w_0+w')}\hspace{0.2cm} exp\Big[-\frac{3w'z}{1+z}\Big],
\end{equation}
\begin{equation}
\frac{\rho_{_{DE}}}{\rho_{_{DE_0}}} = (1+z)^{3(1+w_0)}\hspace{0.2cm} exp\Big[\frac{3w'}{2}\left(\frac{z}{1+z}\right)^2\Big]
\end{equation}
and
\begin{equation}
\frac{\rho_{_{DE}}}{\rho_{_{DE_0}}} = (1+z)^{3(1+w_0)}\hspace{0.2cm}exp\Big[\frac{3w'(z=0)}{2} \left(log(1+z)\right)^2\Big]
\end{equation}
for $p=1$ and $p=2$ and logarithmic parameterisation respectively. Here, $\rho_{_{DE_0}} \equiv
\rho_{_{DE}}(z=0)$, which is the present value of dark energy density,
and $w'  \equiv dw/dz$ at the present time.  

\section{Observations} \label{sec::obsv}

The three data sets we have used in this analysis are the supernovae (SNIa)
observations, Hubble parameter(H(z) observations and the baryon acoustic
oscillation(BAO) data. 
Type Ia supernovae are standard candles and  are useful
in  determining the expansion history of the universe \cite{sn1,sn2,sn3,sn4,sn5,sn6,sn7,sn8,sn9}.   
The supernova's apparent brightness determines its distance from the
observer and the time taken by photons to reach to the observer. 
By comparing the flux to redshift relation for numerous SNIa, the 
rate of expansion of universe and its variation with
time can be determined.
The relation of luminosity distance to redshift $z$ is given by
\begin{equation}
d_L(z) = \frac{c}{H_0} (1+z)\int_0^z d_H(z)dz
\end{equation}
where $d_H$ is the Hubble radius. For a spatially flat universe, 
and if contribution of radiation is
ignored, $\Omega_{m} + \Omega_{DE} = 1$. In this case, $d_H$ is given by
\begin{equation}\label{d_heq}
 d_H(z) = \left[\Omega_{m}(1+z)^3  
+\Omega_{DE} \textrm{exp}\left\{3\int_0^{z} \frac{dz}{1+z}\left[1+w(z)\right]\right\} \right]^{-1/2}.
\end{equation}  

For a $\Lambda$CDM model the equation of state is  $w = -1$.  
The equation can  be further  written as 
\begin{eqnarray}
 d_H(z) &=& \left[\Omega_{m}(1+z)^3 +\Omega_{DE} \right]^{-1/2}.
\end{eqnarray}
For a $w$CDM model, $w\neq-1$ and is a constant.
The Eq. \ref{d_heq} then modifies as
\begin{eqnarray}
 d_H(z) &=& \left[\Omega_{m}(1+z)^3 +\Omega_{DE}(1+z)^{3(1+w)} \right]^{-1/2}.
\end{eqnarray}

The SNIa data comprises of distance modulus at a 
given redshift along with the associated error
\cite{sn9}.
The distance  modulus, $\mu$, is defined as
\begin{equation}
\mu = 5~\textrm{log}(d_L) - 5,
\end{equation}
in $10$ $pc$ units.
\begin{table}
\centering
\begin{tabular}{|c|c|c|} \hline
Parameter & Lower Limit & Upper Limit\\ \hline
$\Omega_{m}$&0.01 & 0.6\\ \hline
w& -4.0 & 0.0\\ \hline
$H_0$&65.0 &75.0\\ \hline
\end{tabular}

\caption{\label{table::priors} This table lists the priors used for parameter fitting in case of $w$CDM model.}
\end{table}

%\subsection{BAO Data}
The second dataset we have used is that of Baryon Acoustic Oscillations
\cite{bao00,bao1, bao2, bao3,bao4,new_bao1, new_bao2}. 
Before the recombination epoch, baryons are  tightly coupled to
photons via the Thompson scattering.  
The competition between the  pressure and gravity leads 
to acoustic oscillations.
These sound waves are imprinted in the baryon perturbations, and give
rise to baryon acoustic oscillation peaks.   
These constitute  an independent  probe of  dark energy as they
provide a standard ruler for length scale in cosmology, namely the
radius of the sound horizon.
The  effective distance ratio  is defined as 
\begin{equation}
 D_v(z)= \Big[\frac{(1+z)^2d_A^2(z)cz}{H(z)}\Big]^{1/3},
\end{equation}  
where $d_A$ is the angular diameter distance. 
Another description  has been introduced by Eisenstein \cite{bao00}, where
the acoustic parameter is defined as   
\begin{equation}
 A(z)= \frac{100D_v(z)\sqrt{(\Omega_{m}h^2)}}{cz}.
\end{equation} 
The BAO data comprises of $A(z)$ and $d_z(z)$ along with errors associated
with them for different values of $z$ \cite{bao00,bao1, bao2, bao3,bao4,new_bao1, new_bao2}.
For our analysis, we have used the Acoustic Parameter for six data points
from \cite{bao4} and the  parameter $D_V(z)$ for the three  data
points in \cite{new_bao1,new_bao2}. 
 
\begin{table}
  \centering
\begin{tabular}{|c|c|c|} \hline
Parameter & Lower Limit & Upper Limit\\ \hline
$\Omega_{m}$& 0.1 & 0.6\\ \hline
$w_0$& -5.0 & 2.0\\ \hline
$w'(z=0)$&-10.0& 10.0\\ \hline
$H_0$&65.0 &75.0\\ \hline
\end{tabular}
\caption{\label{table::priors_two} This table lists the priors used in the parameter fitting, 
when $w$ is a function of redshift.}
\end{table}

%\subsection{Hubble Parameter Measurement (H(z)) data}
The value of Hubble parameter at different redshifts is also a
useful tool to constrain the cosmological parameters \cite{hz1,hz2,hz3,hz4,hz5,hz6,hz7,hz8,hz9,hz10}.
The Hubble parameter H(z) is measured using different techniques. 
The data comprises of $28$ independent measurements of H(z), 
listed in  references in \cite{hz1}.

\begin{figure}
\centering
%\begin{tabular}{cccc}
\includegraphics[width=0.2\textwidth]{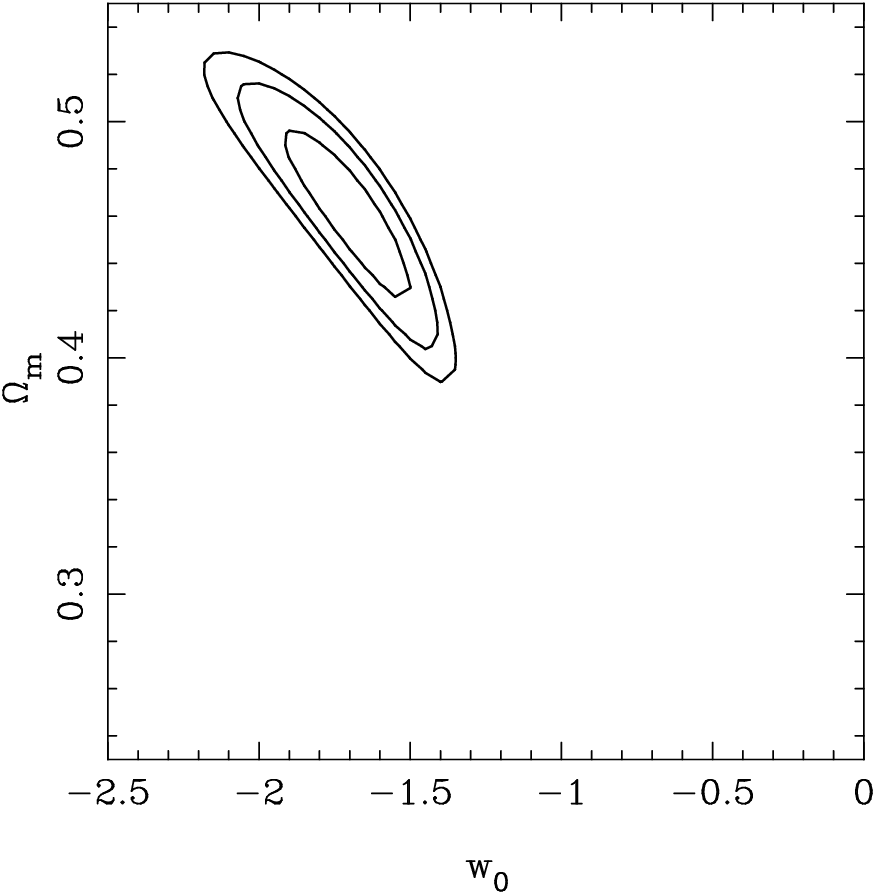} \includegraphics[width=0.2\textwidth]{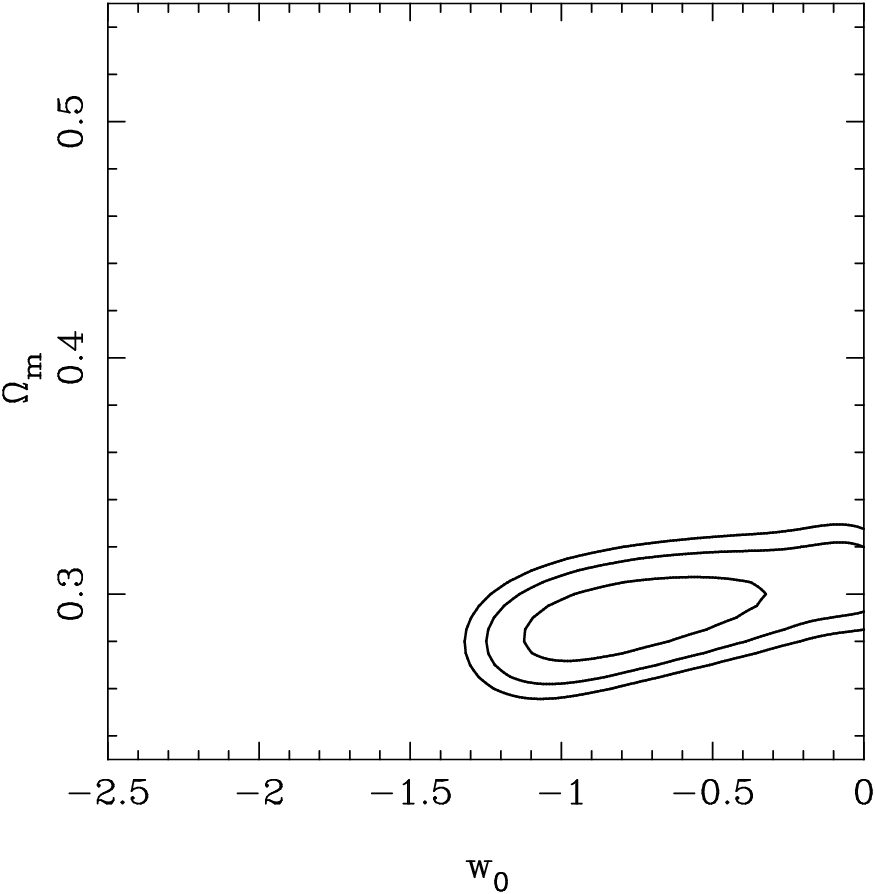}\includegraphics[width=0.2\textwidth]{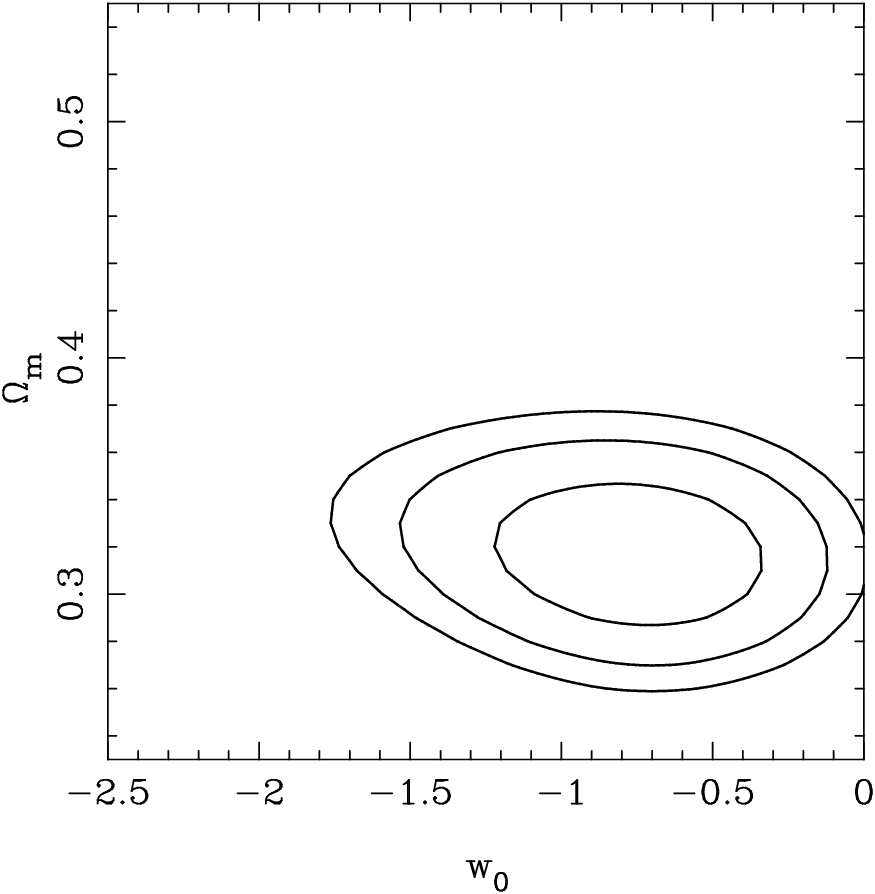}\includegraphics[width=0.2\textwidth]{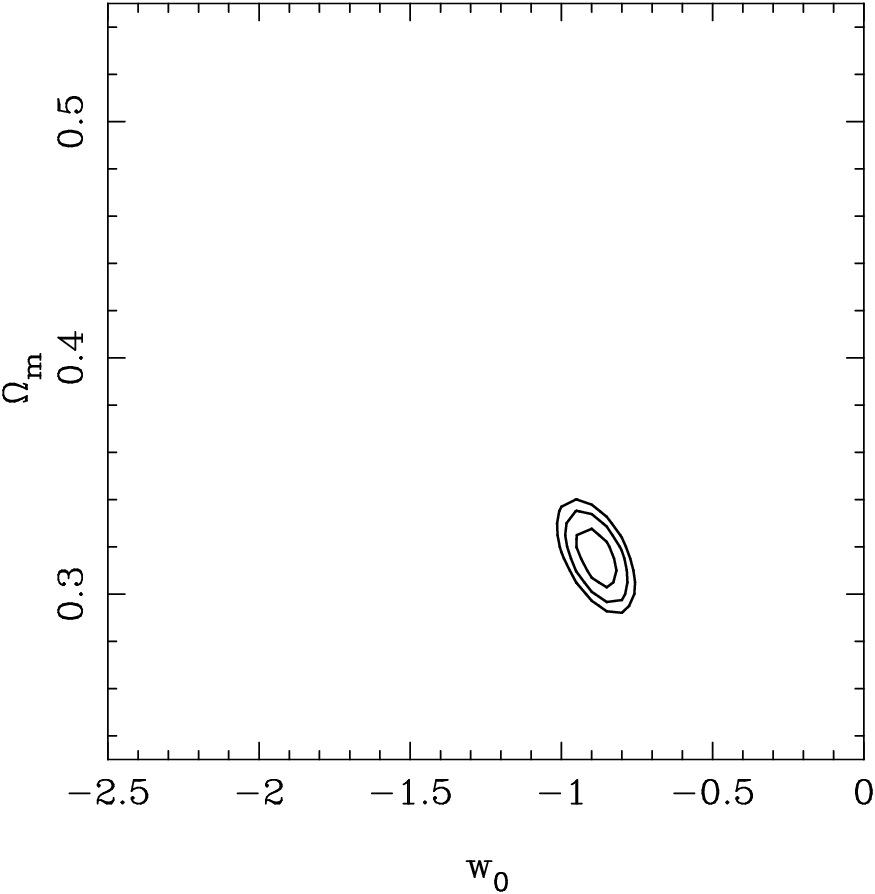}\\
\includegraphics[width=0.2\textwidth]{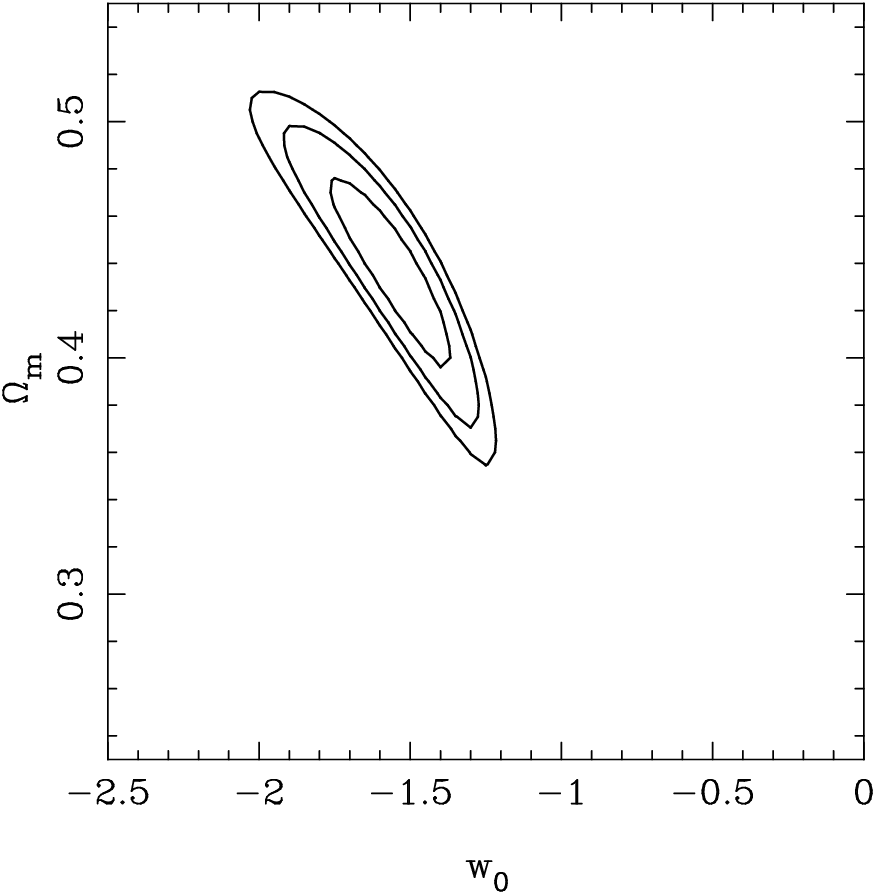}\includegraphics[width=0.2\textwidth]{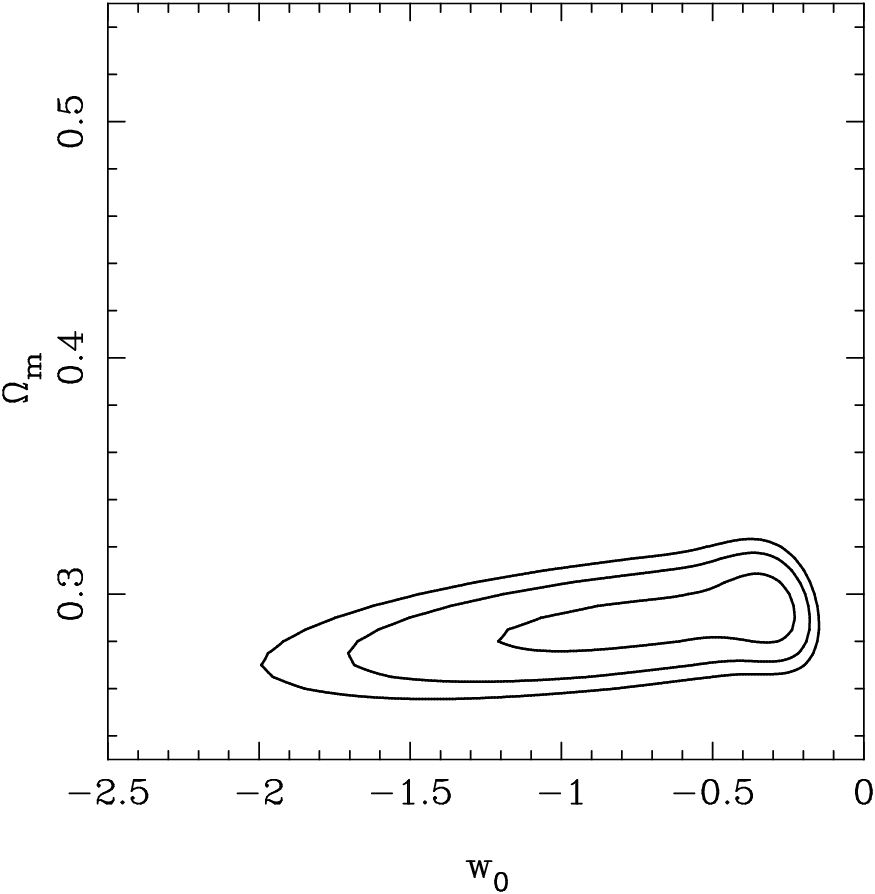}\includegraphics[width=0.2\textwidth]{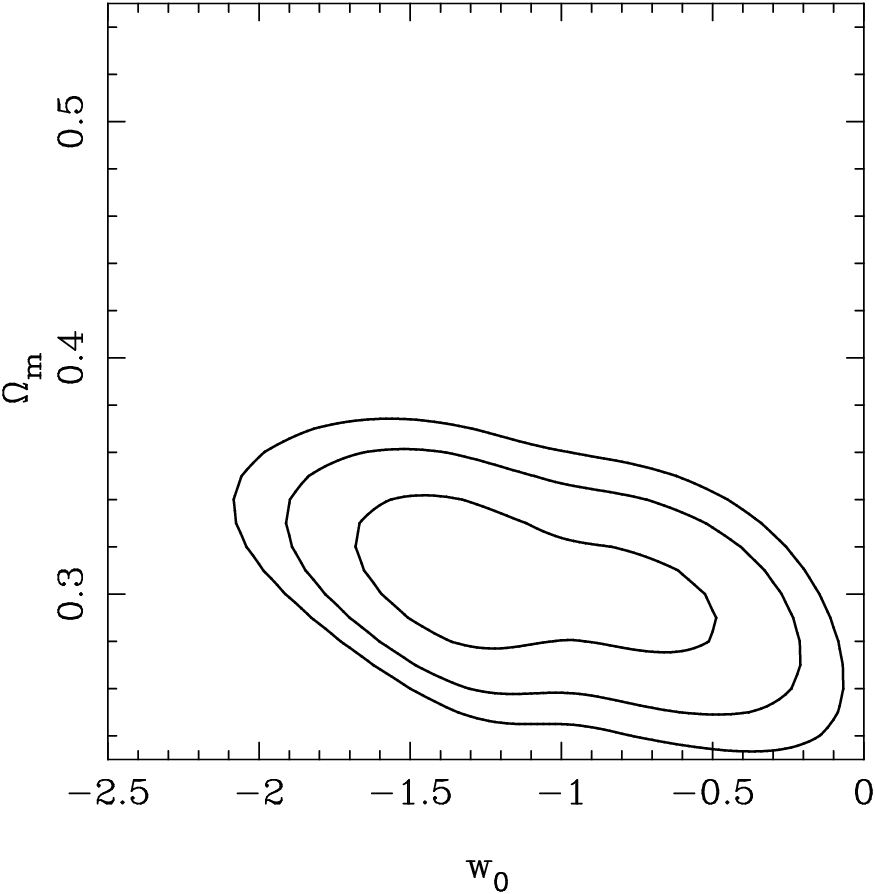}\includegraphics[width=0.2\textwidth]{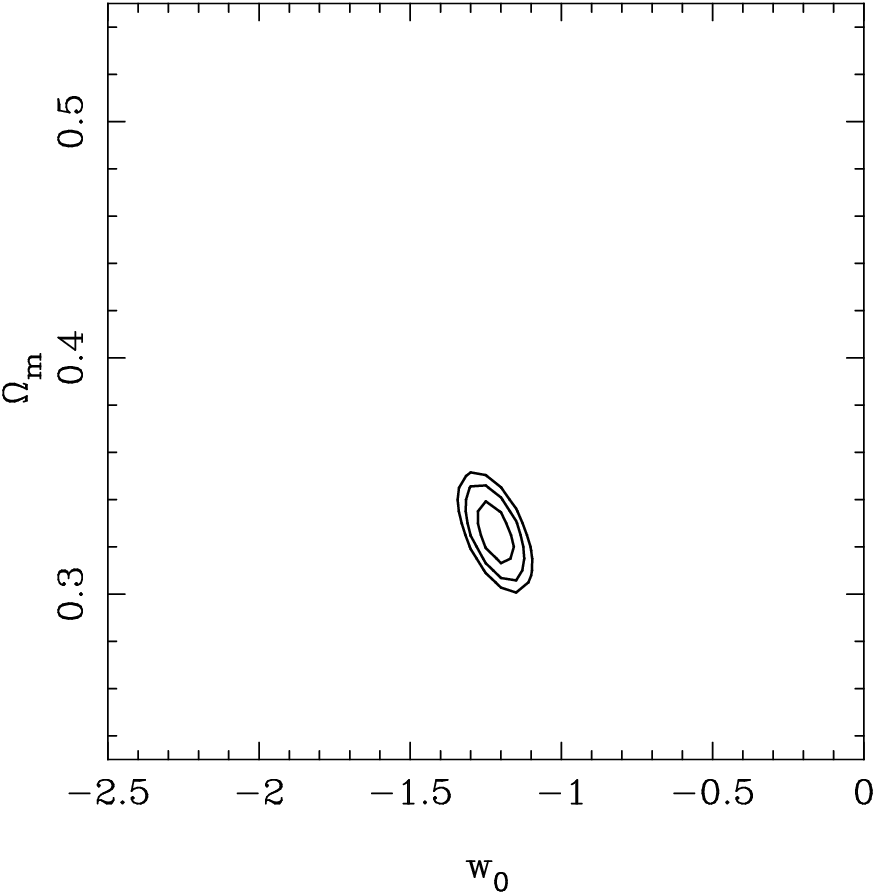}\\
\includegraphics[width=0.2\textwidth]{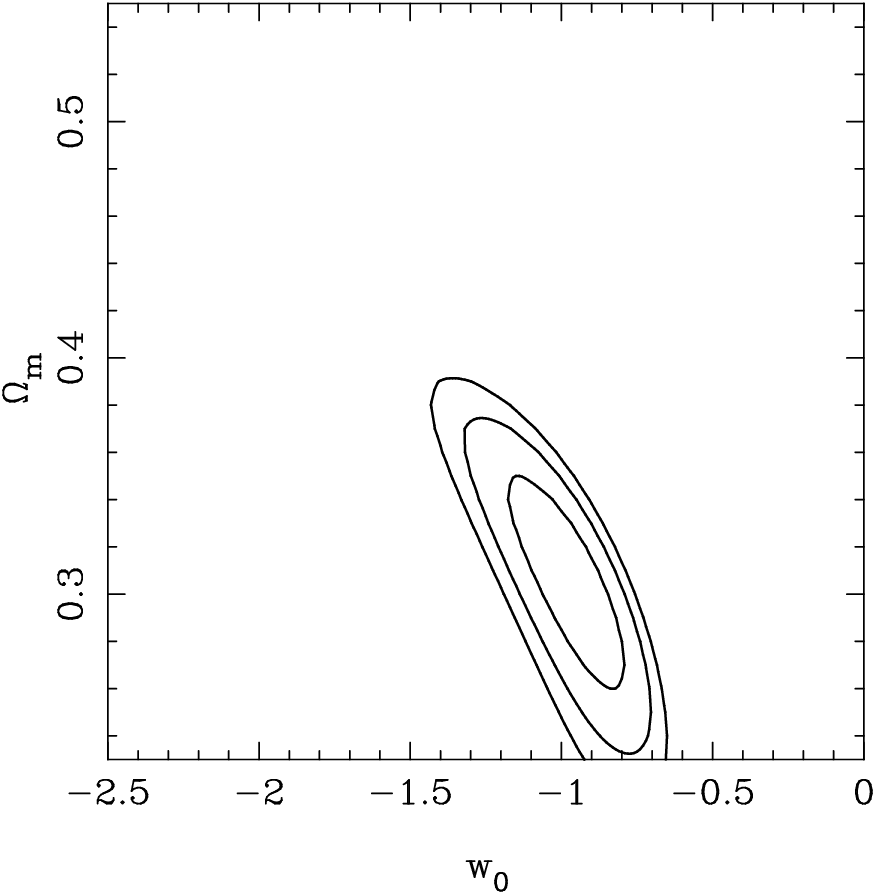}\includegraphics[width=0.2\textwidth]{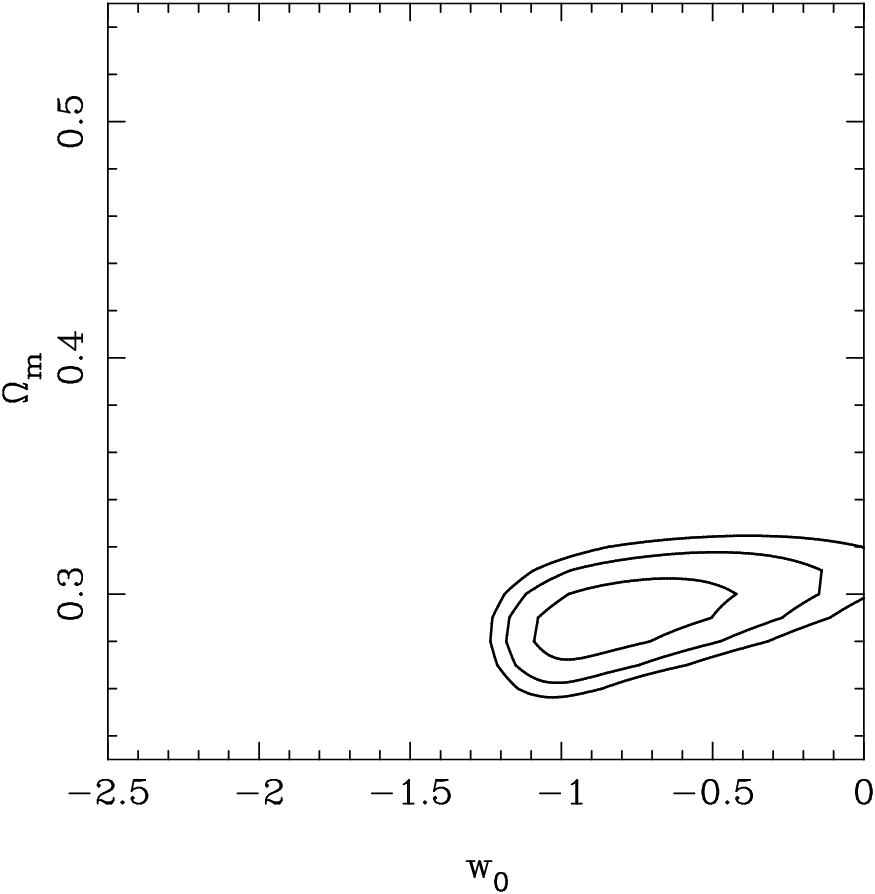}\includegraphics[width=0.2\textwidth]{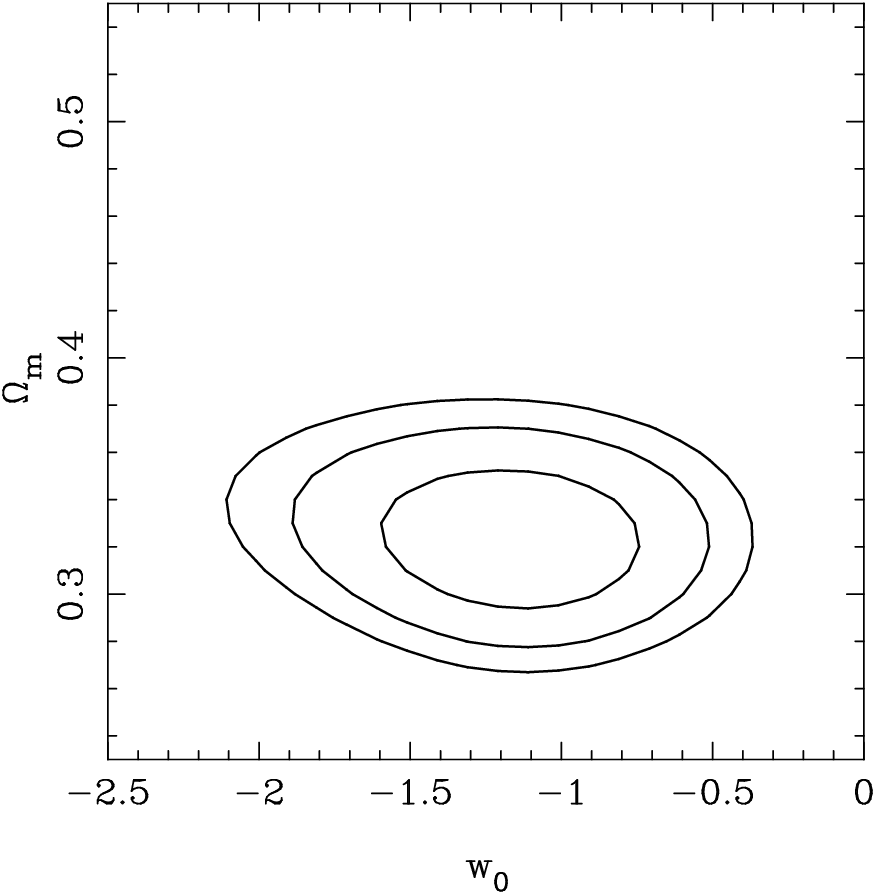}\includegraphics[width=0.2\textwidth]{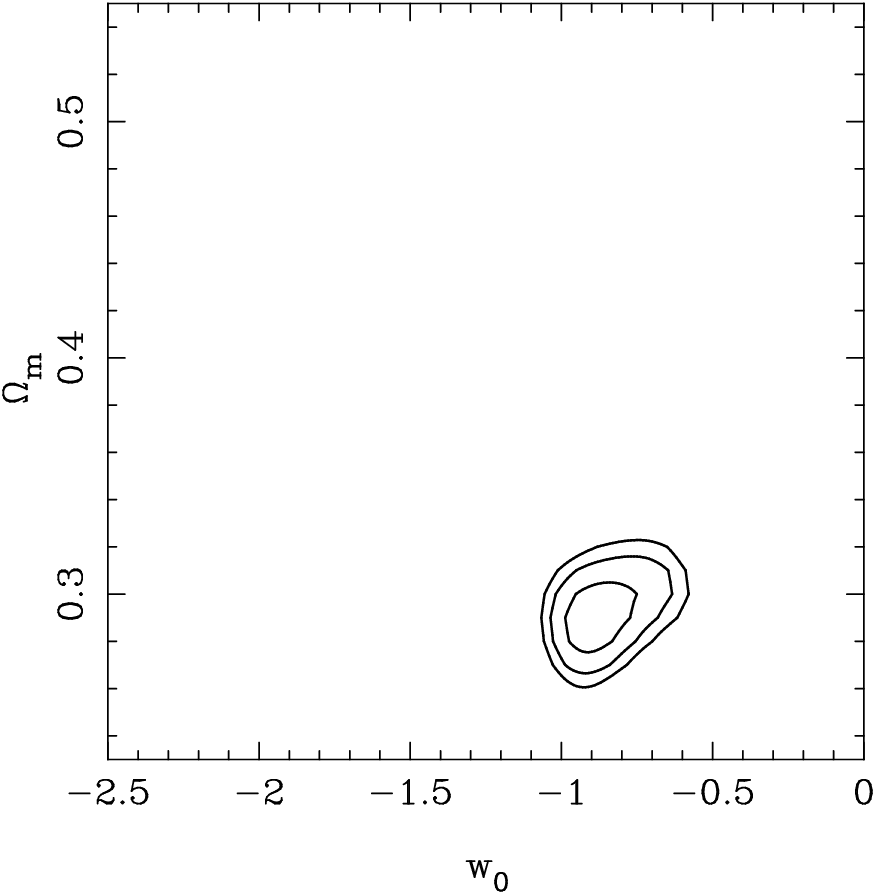}\\
%\end{tabular}
\caption{\label{fig::marginpar}   The plots 
  represent the confidence contours for (from
  left to right) SNIa, BAO, H(z) and a combination of the datasets for
  the two different parameterizations with marginalization over $w'$. 
  The contours in first and second row are obtained for
  parameterizations $p=1$ and 
  $p=2$ respectively and plots in the last row are for the logarithmic paramterization.
 } 

\end{figure}

\begin{figure}
\centering
%\begin{tabular}{cccc}
\includegraphics[width=0.2\textwidth]{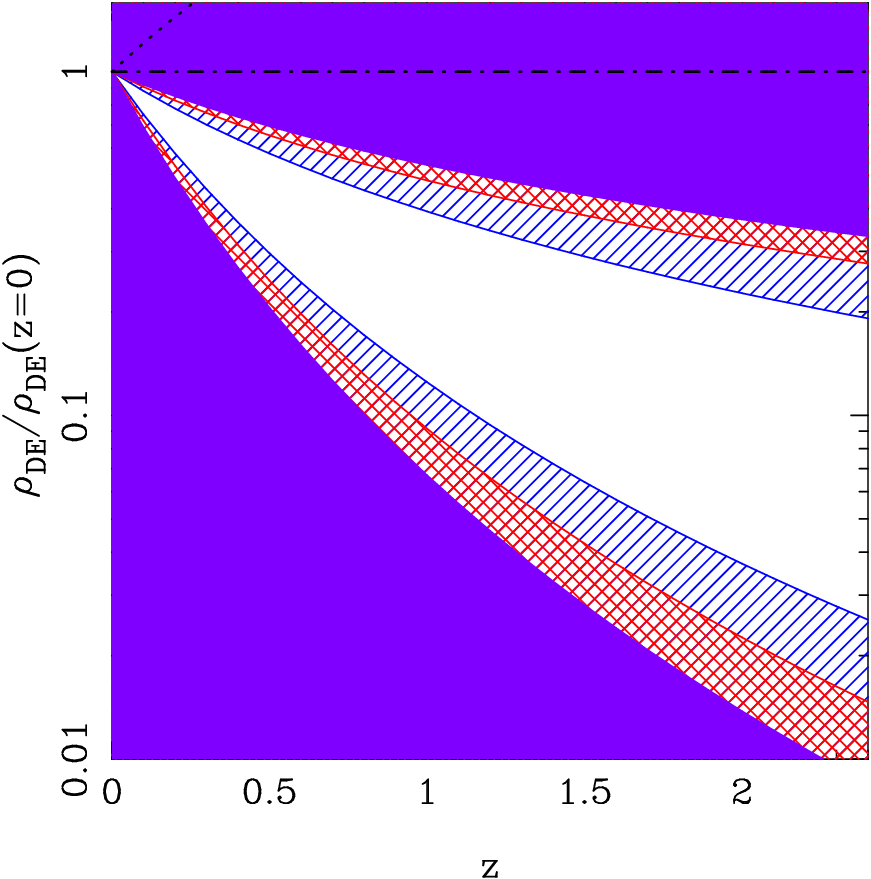}\includegraphics[width=0.2\textwidth]{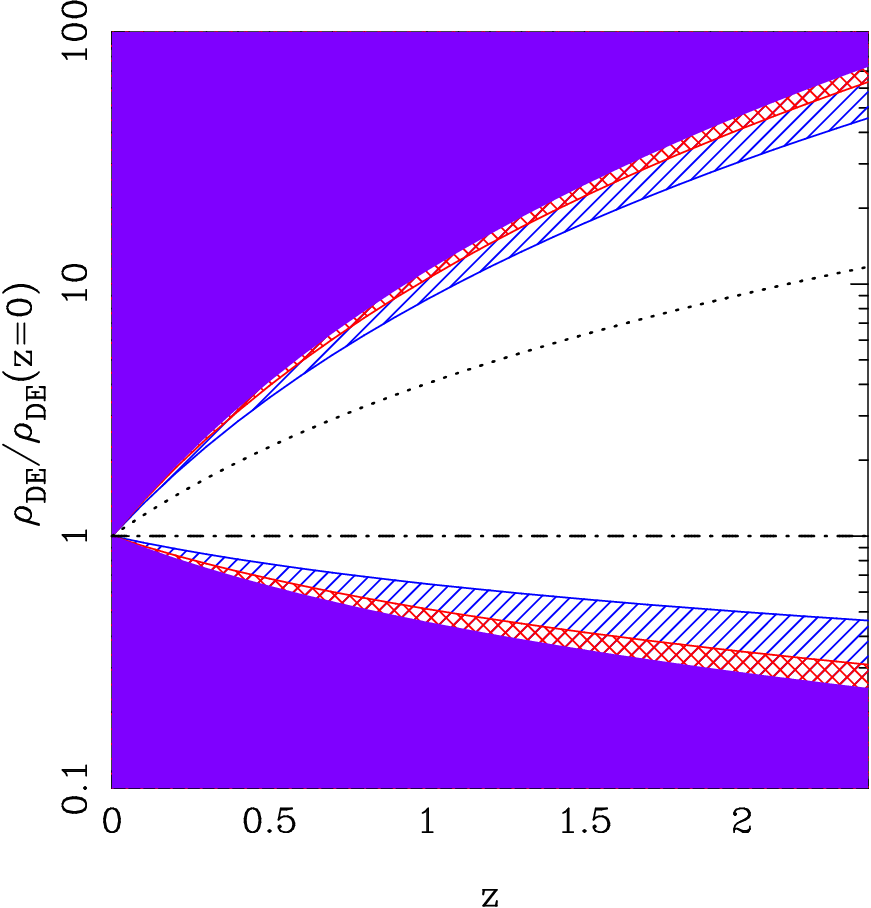}\includegraphics[width=0.2\textwidth]{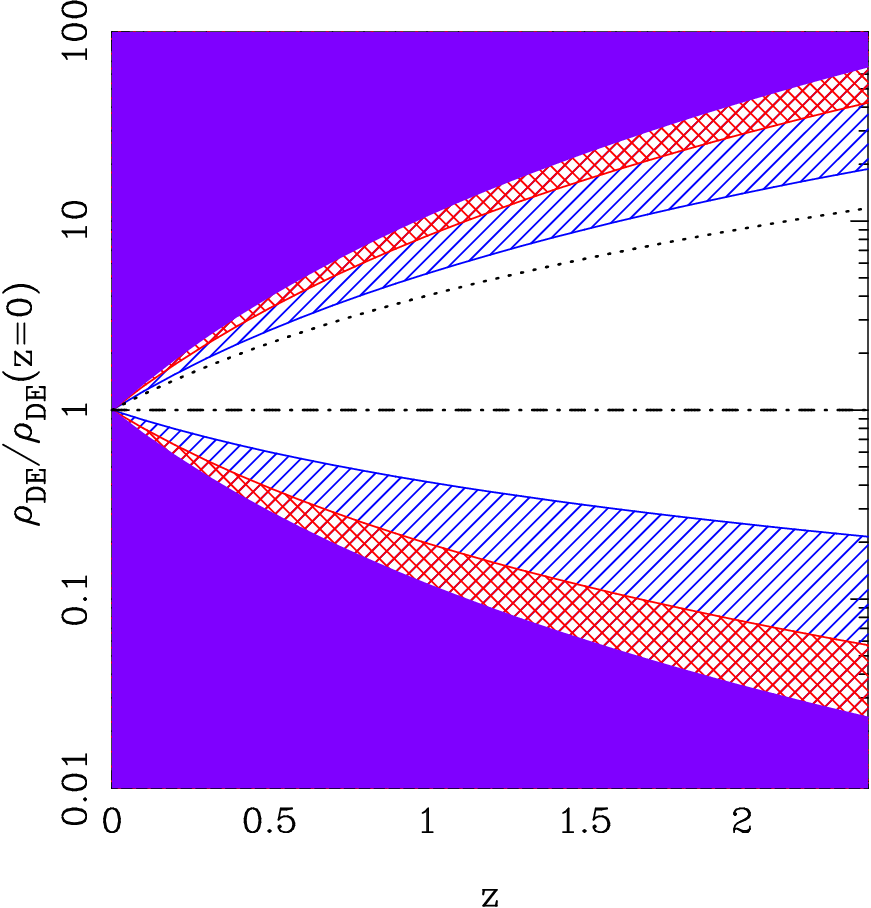}\includegraphics[width=0.2\textwidth]{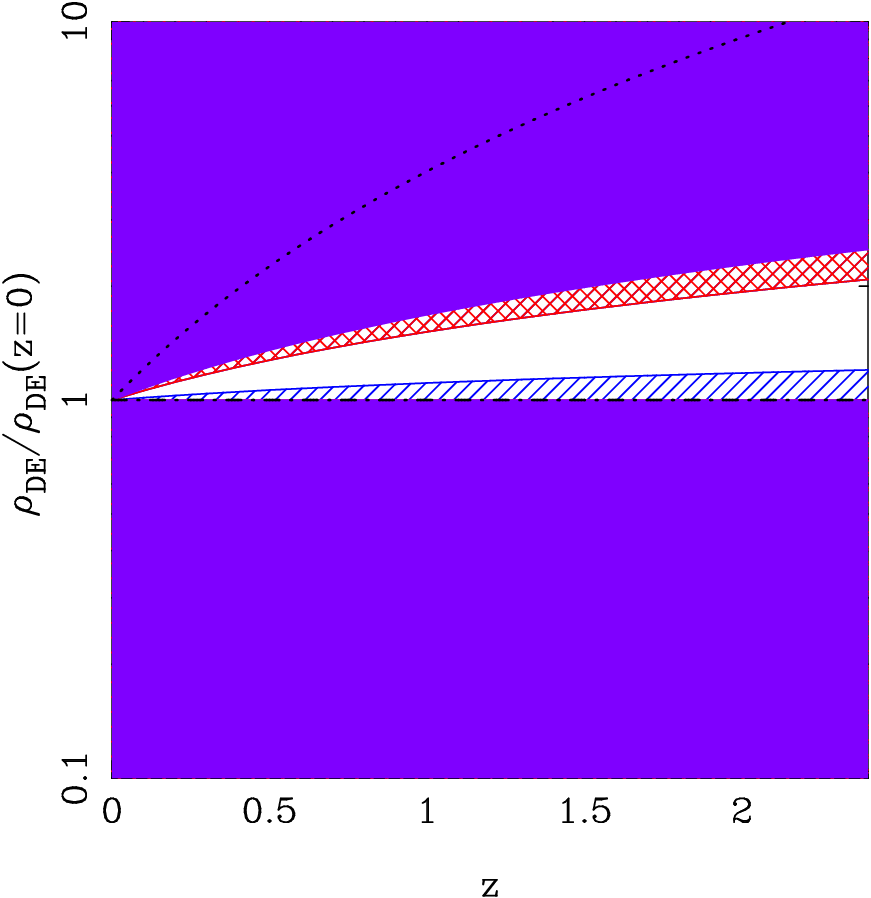}\\
\includegraphics[width=0.2\textwidth]{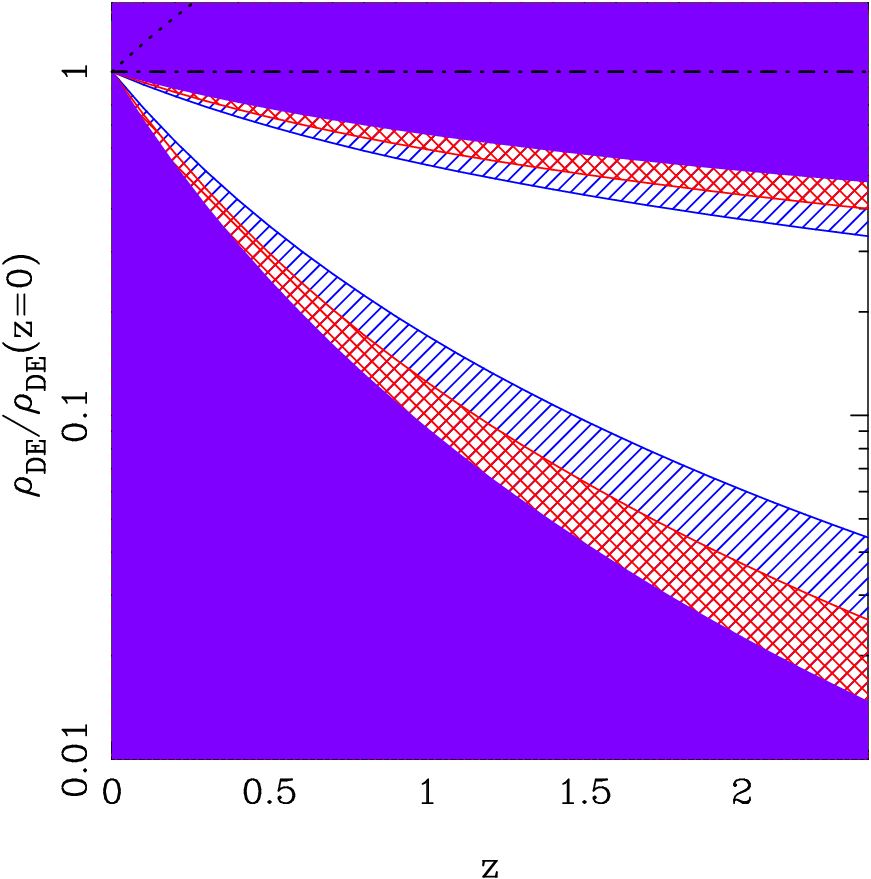}\includegraphics[width=0.2\textwidth]{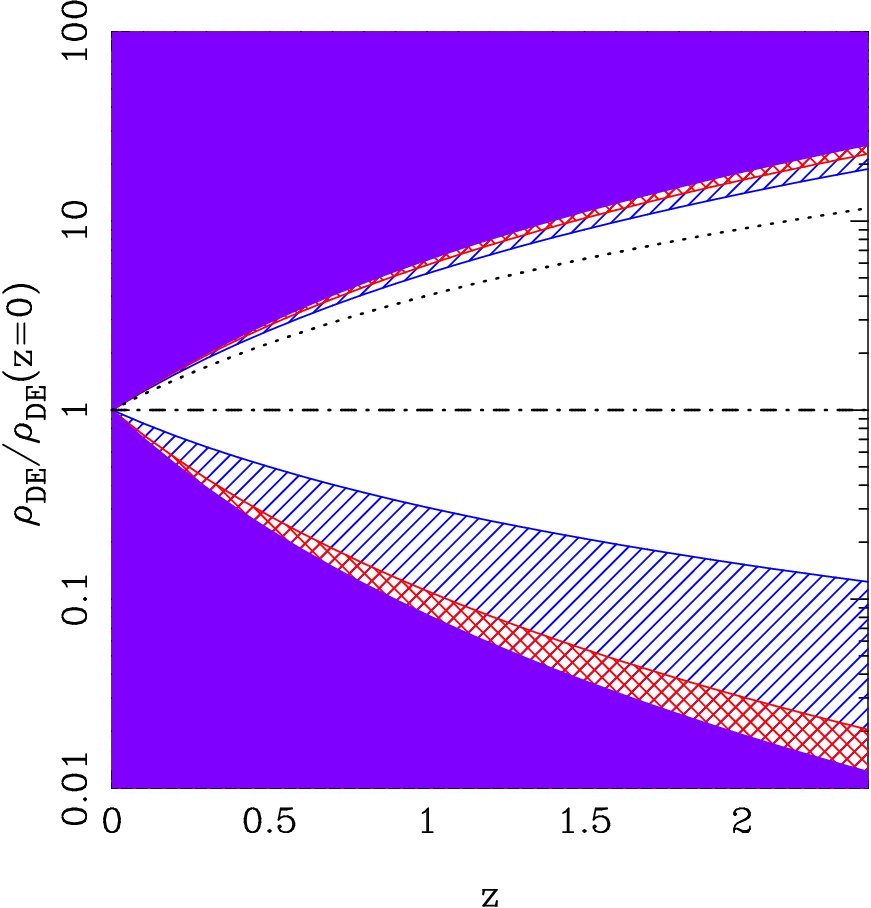}\includegraphics[width=0.2\textwidth]{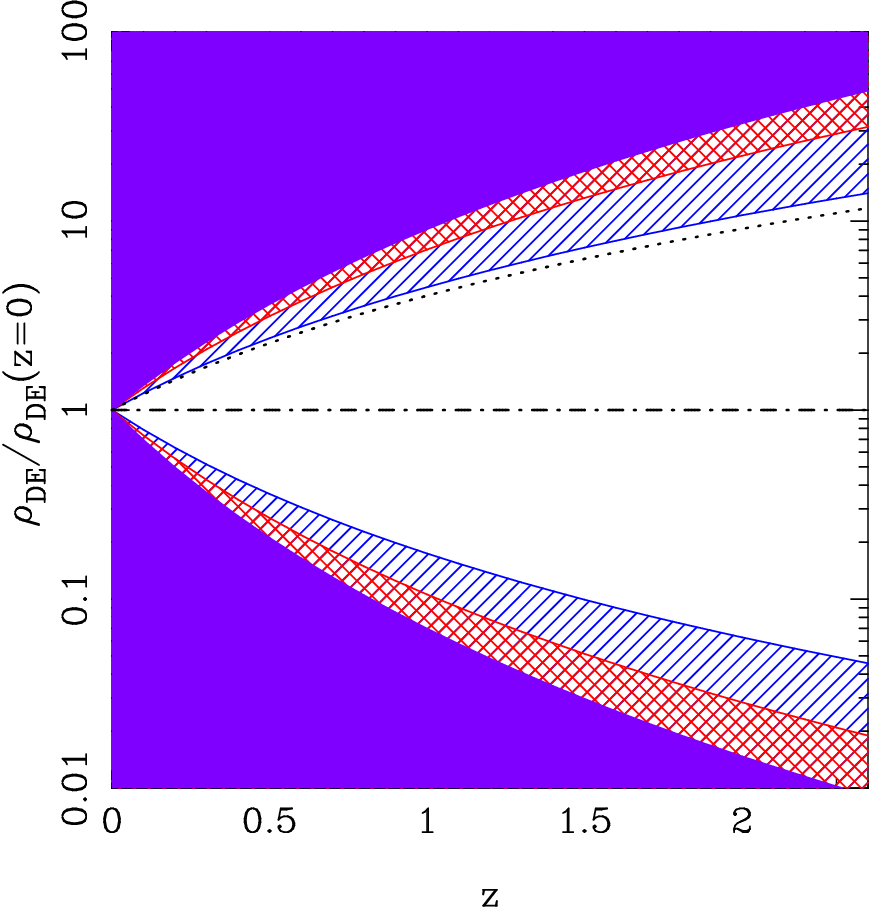}\includegraphics[width=0.2\textwidth]{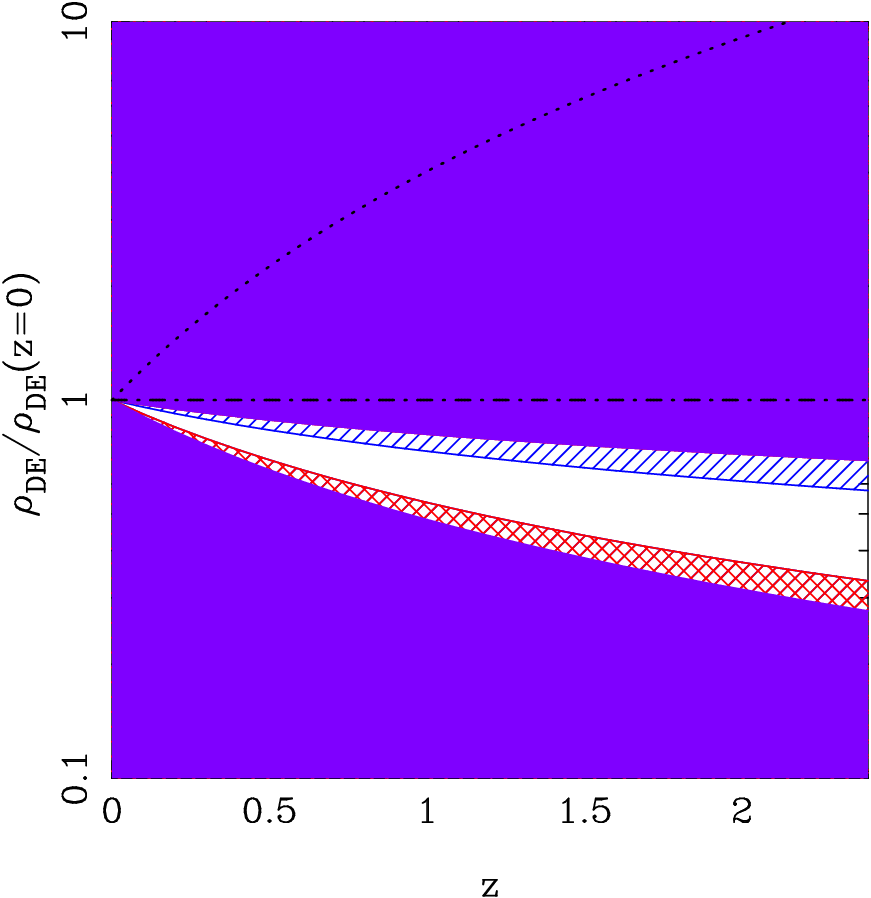}\\
\includegraphics[width=0.2\textwidth]{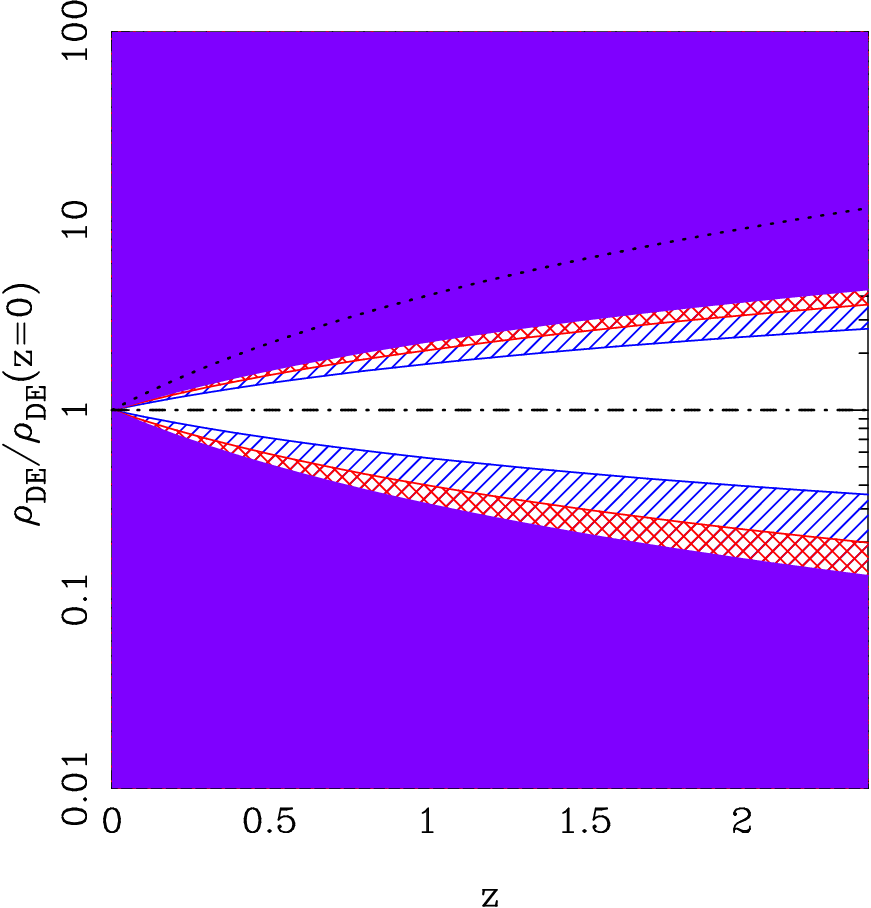}\includegraphics[width=0.2\textwidth]{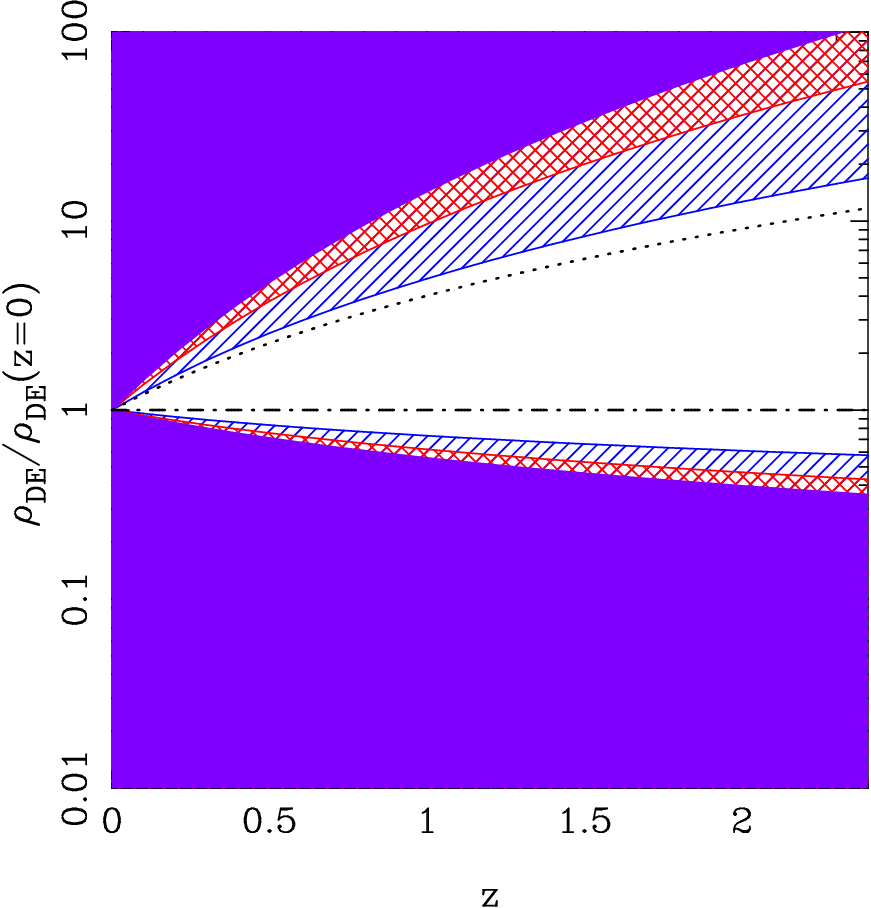}\includegraphics[width=0.2\textwidth]{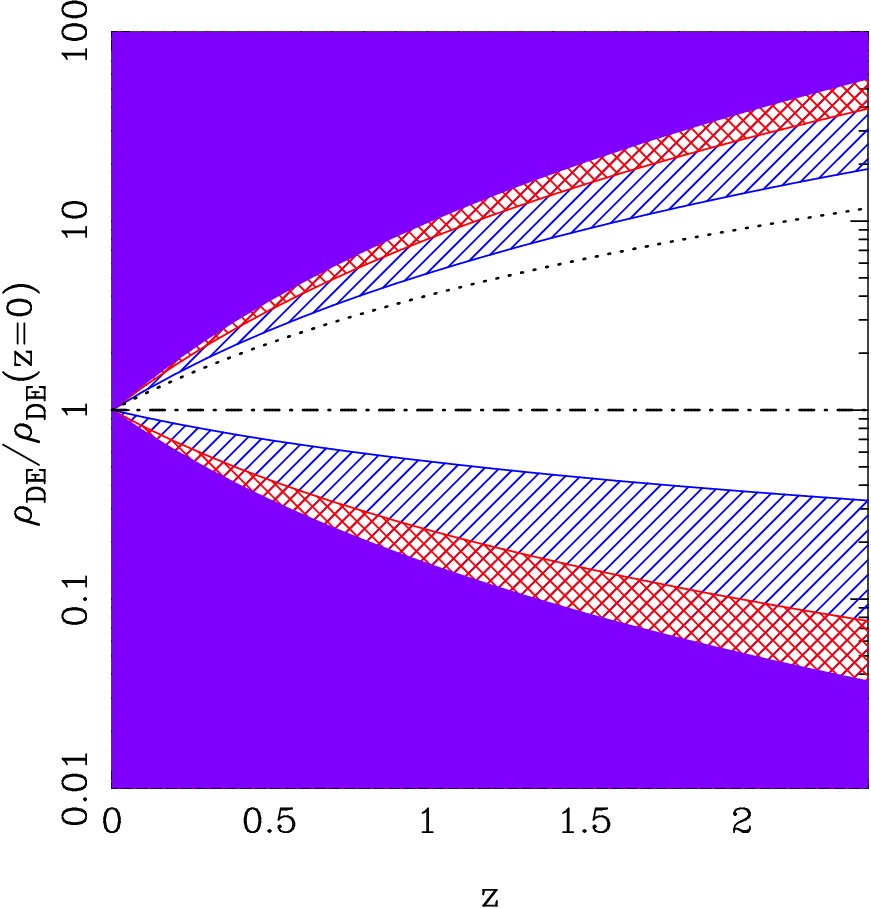}\includegraphics[width=0.2\textwidth]{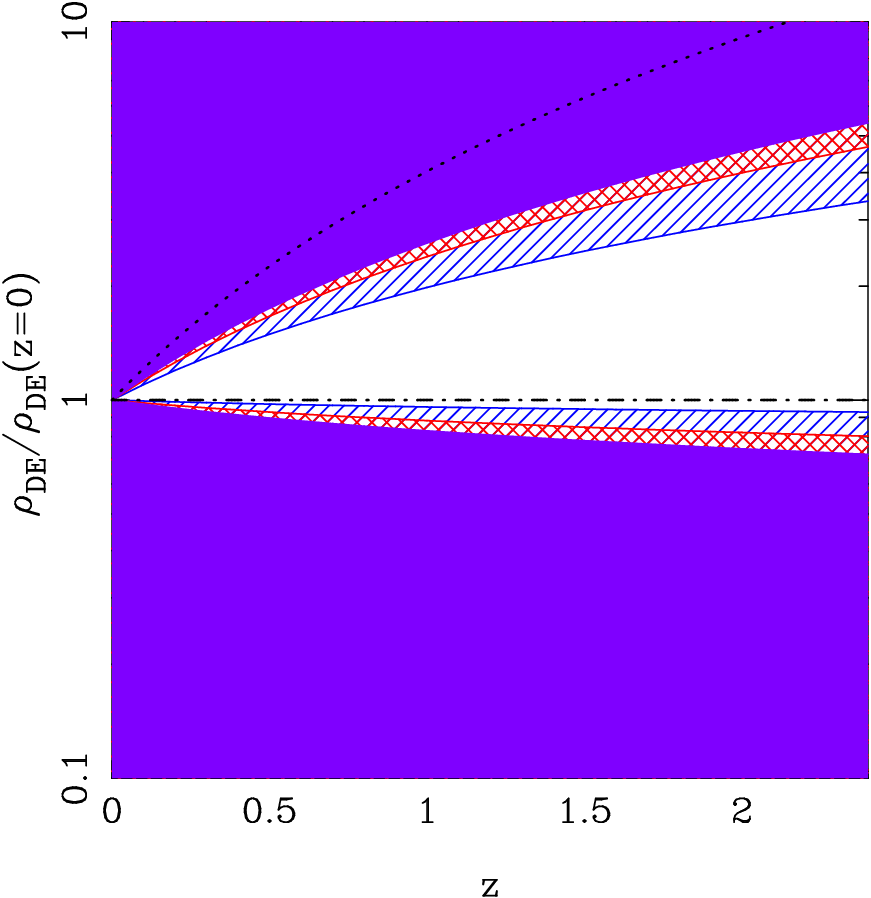}\\

%\end{tabular}
\caption{\label{fig::marginpar1}   The plots 
  represent the variation of dark energy density as a function of redshift allowed at 1, 2 and 3 -$\sigma$ confidence levels. From  left to right the ranges allowed by SNIa, BAO, H(z) and a combination of the datasets for  
  the two different parameterizations with marginalization over $w'$. 
  The three rows correspond the models in \ref{fig::marginpar} } 

\end{figure}

\begin{figure}
\centering
%\begin{tabular}{cccc}
\includegraphics[width=0.2\textwidth]{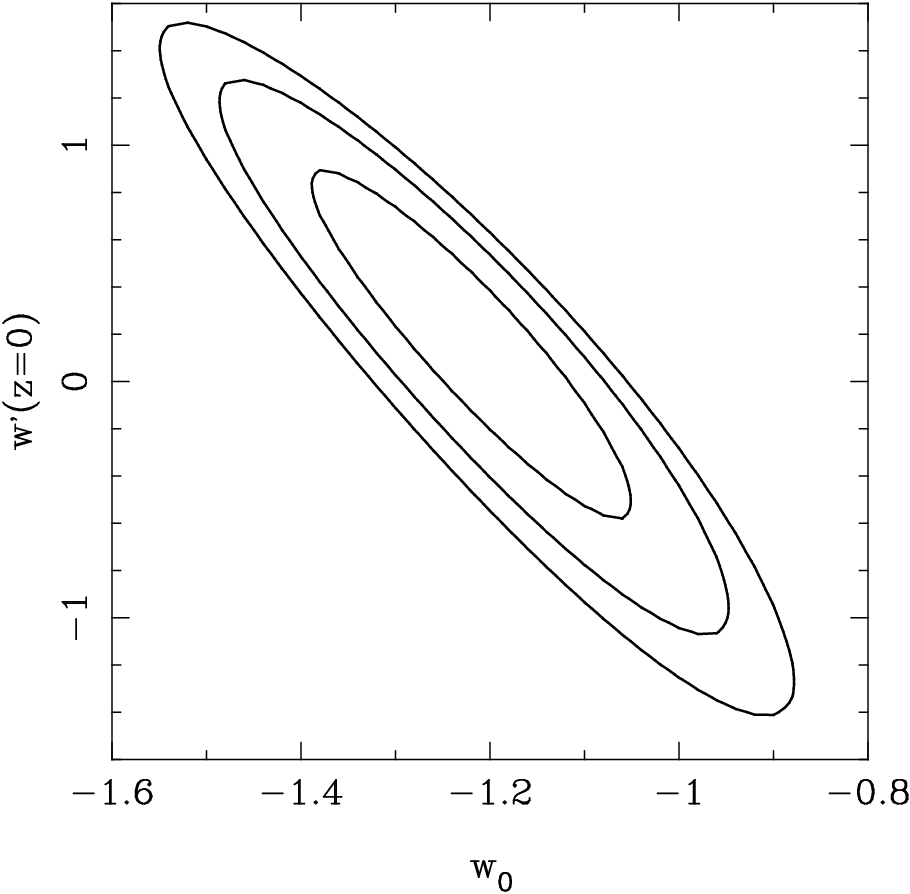}\includegraphics[width=0.2\textwidth]{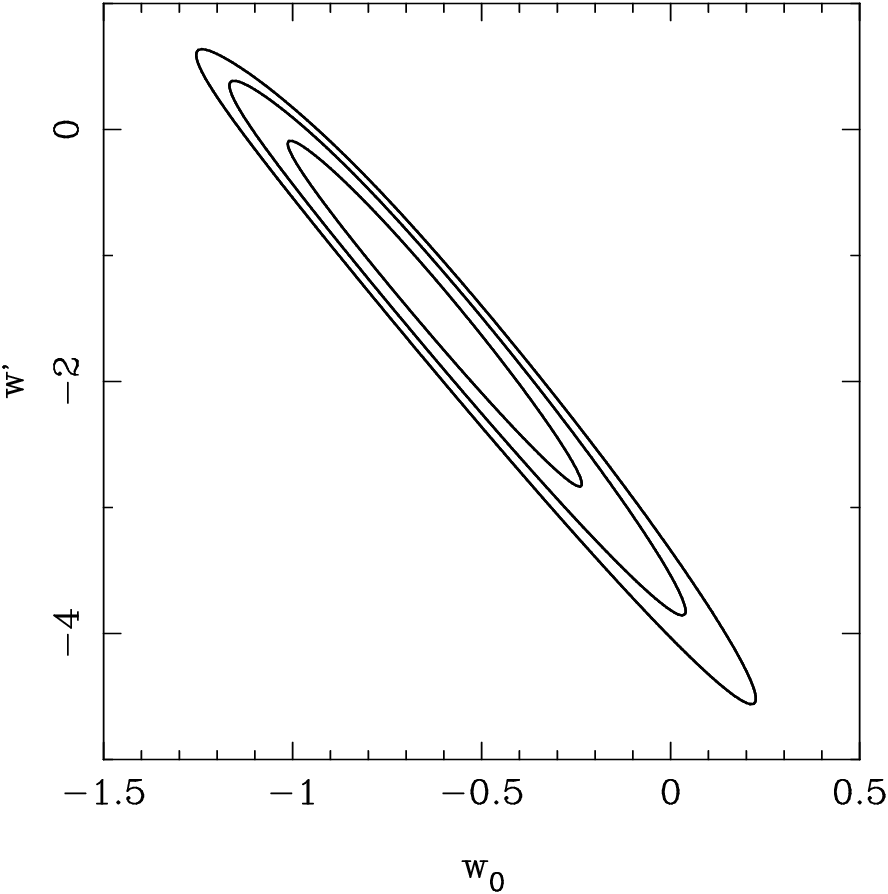}\includegraphics[width=0.2\textwidth]{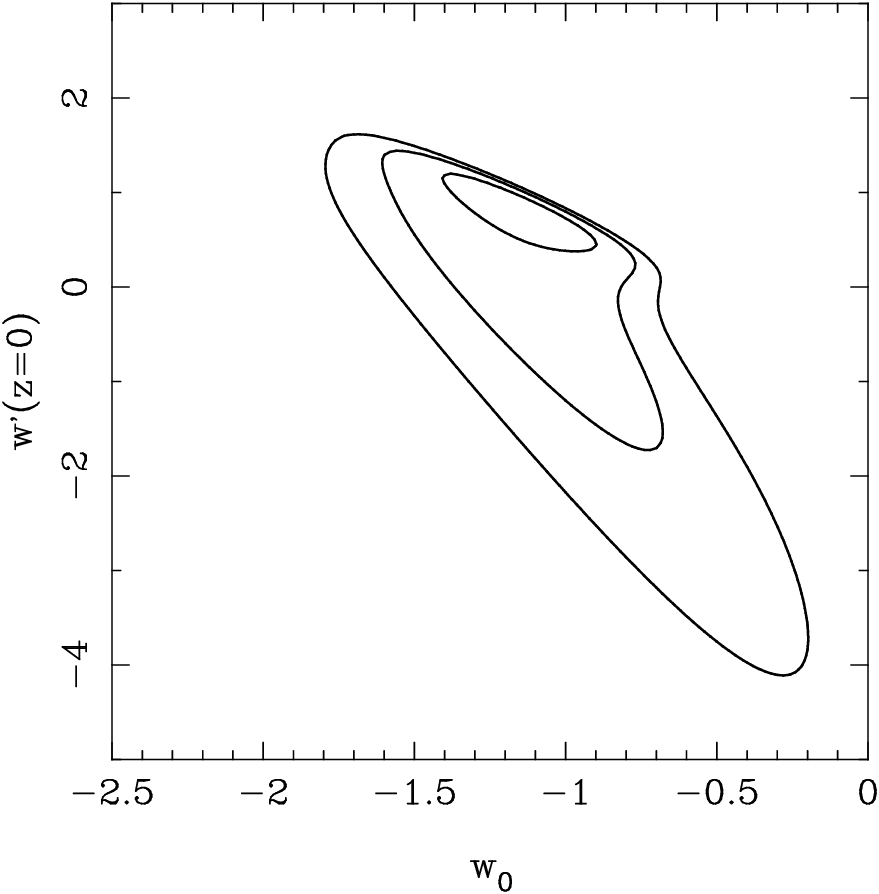}\includegraphics[width=0.2\textwidth]{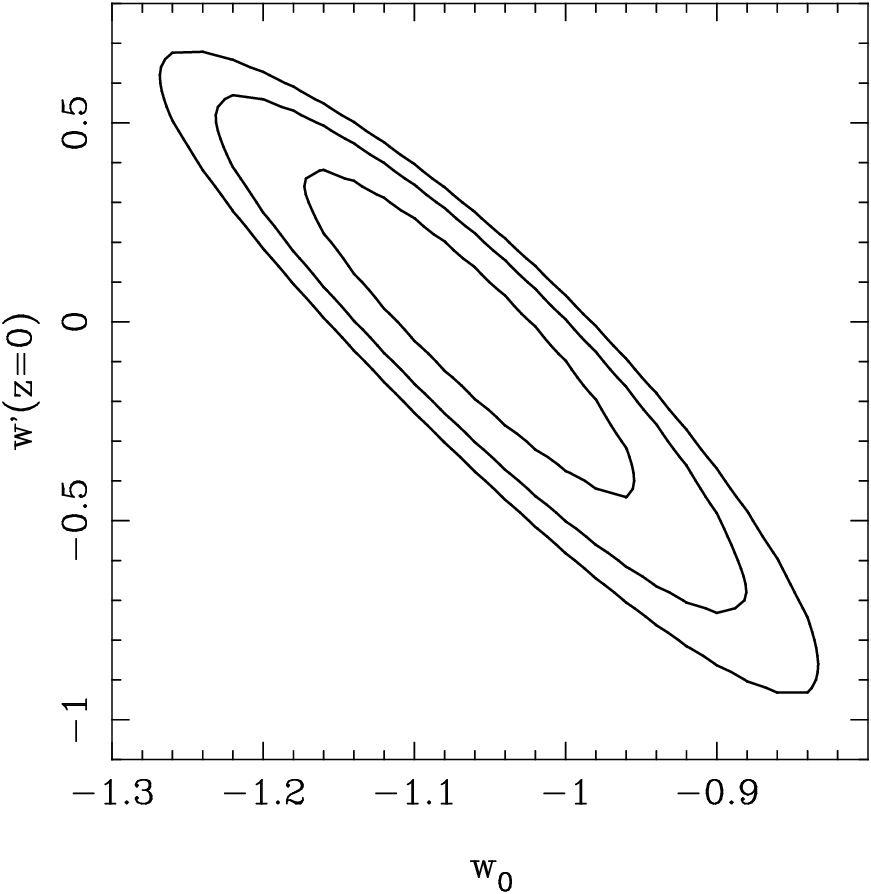}\\
\includegraphics[width=0.2\textwidth]{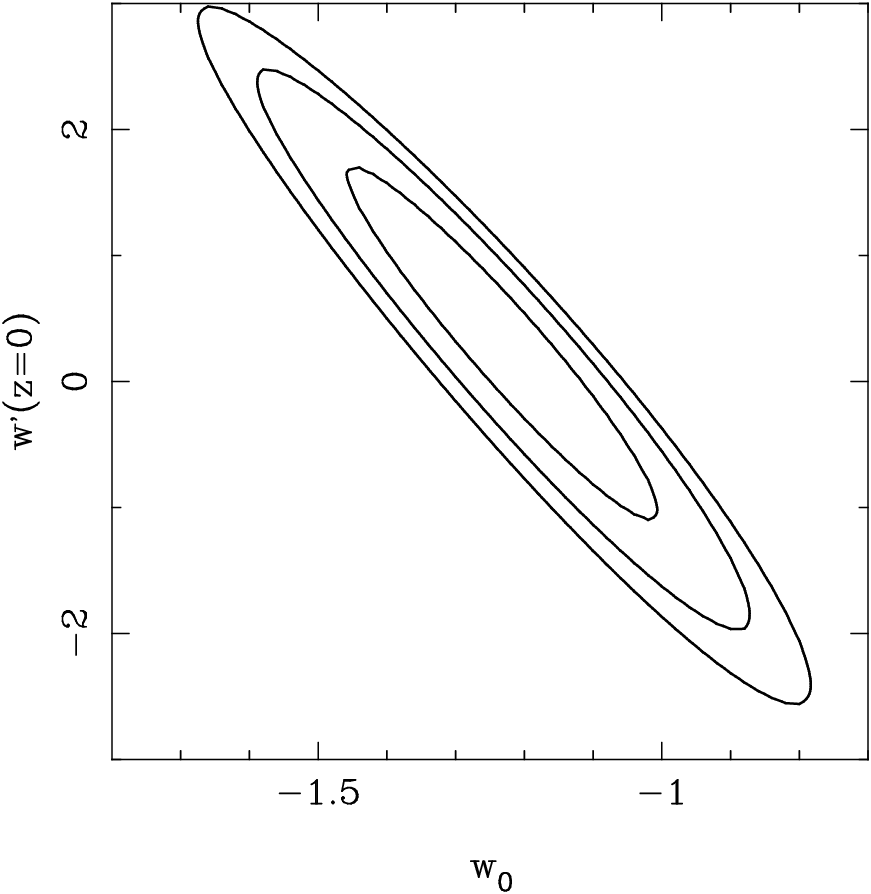}\includegraphics[width=0.2\textwidth]{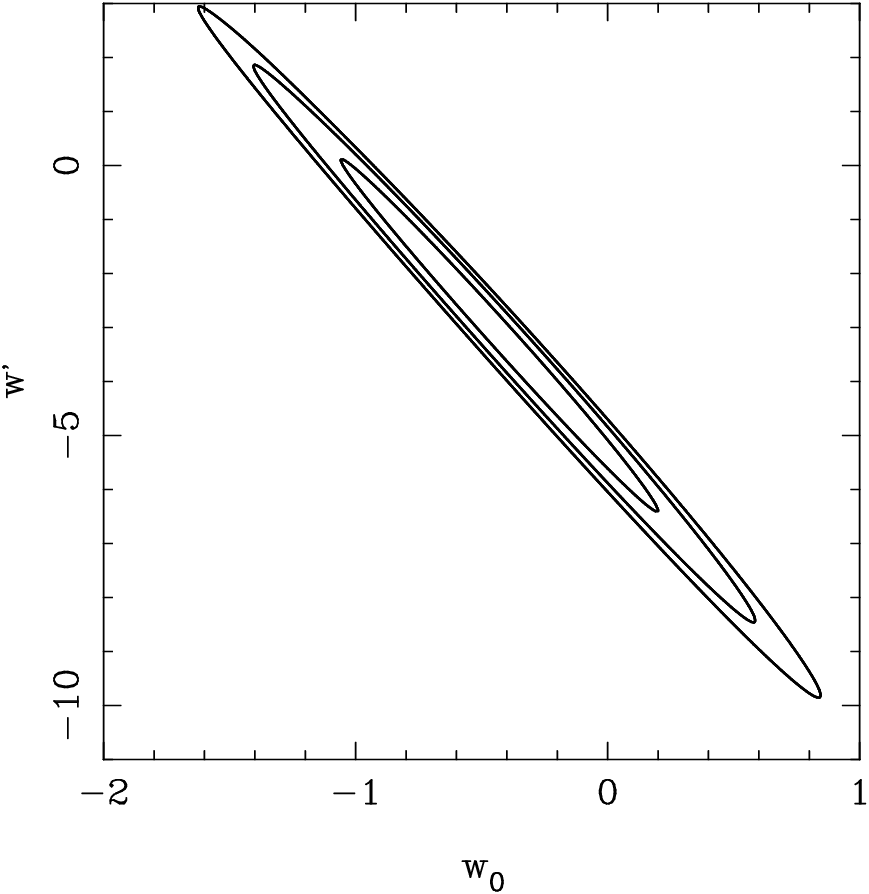}\includegraphics[width=0.2\textwidth]{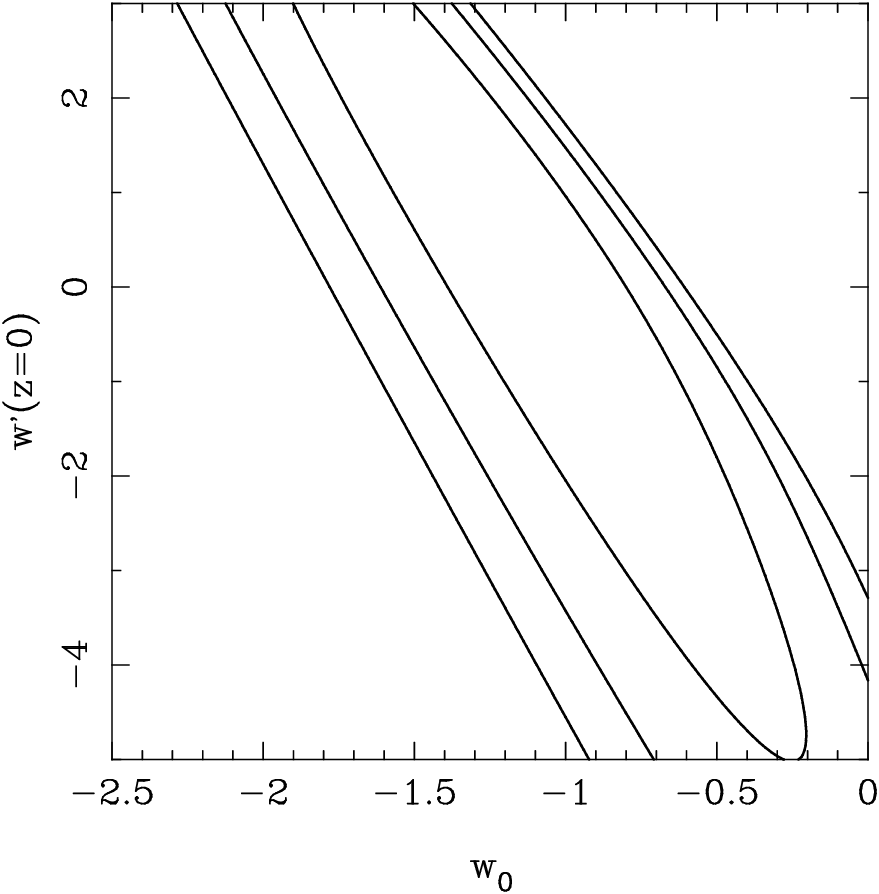}\includegraphics[width=0.2\textwidth]{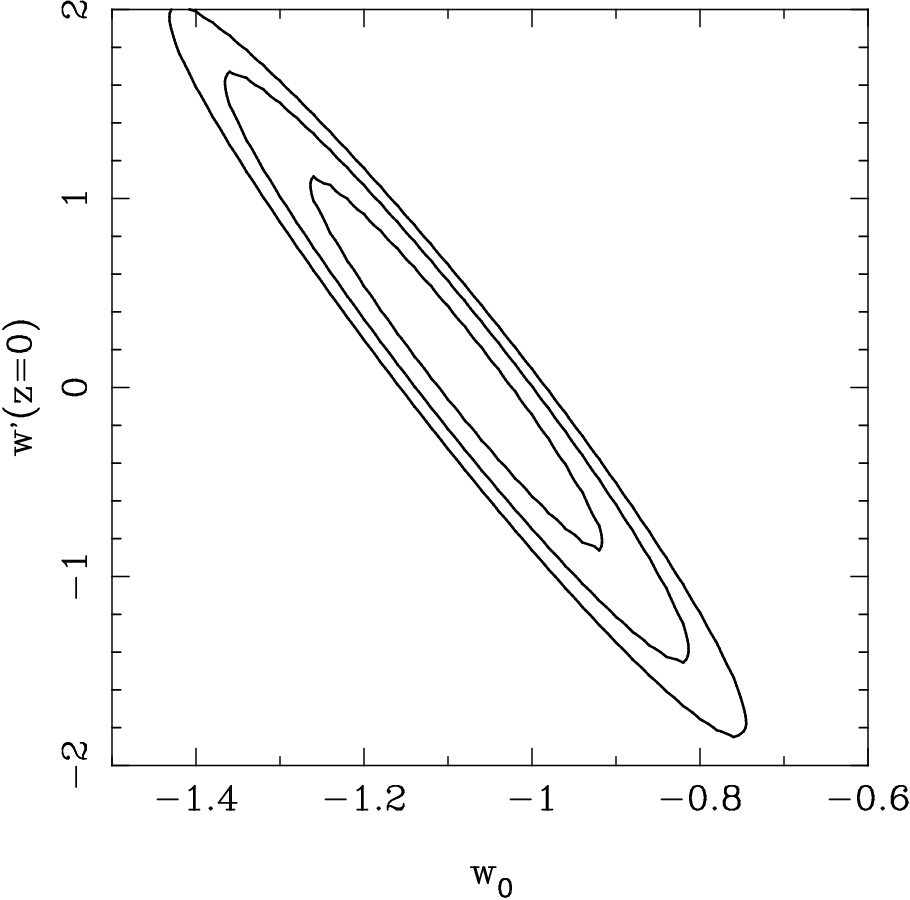}\\
\includegraphics[width=0.2\textwidth]{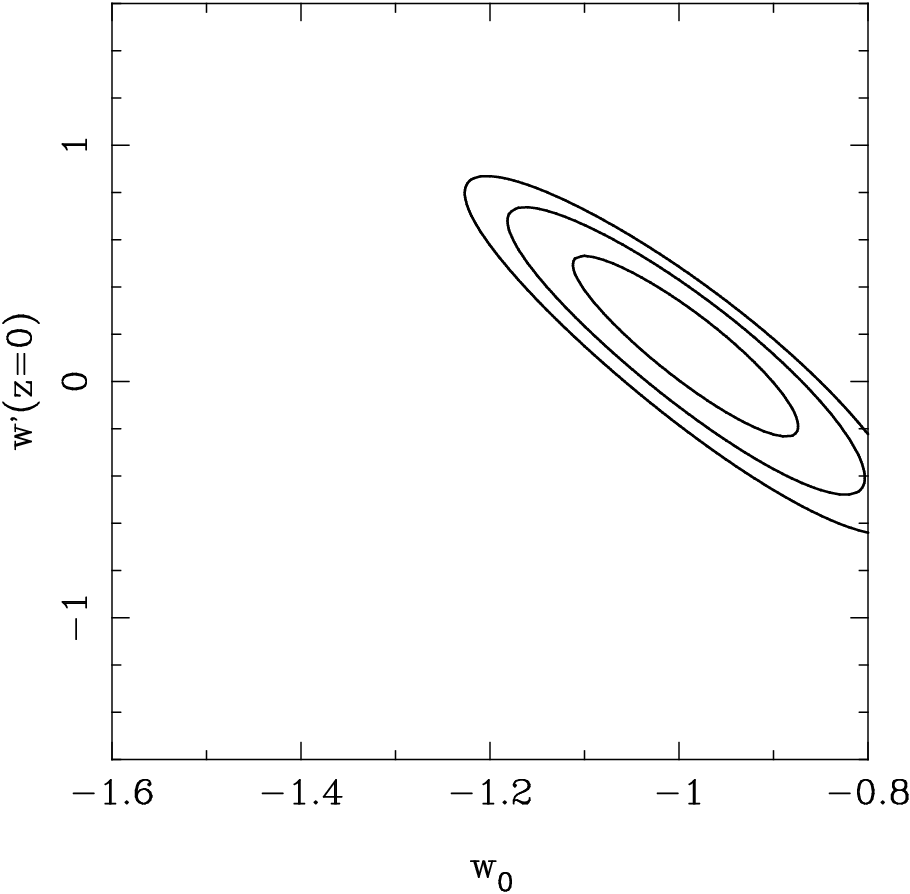}\includegraphics[width=0.2\textwidth]{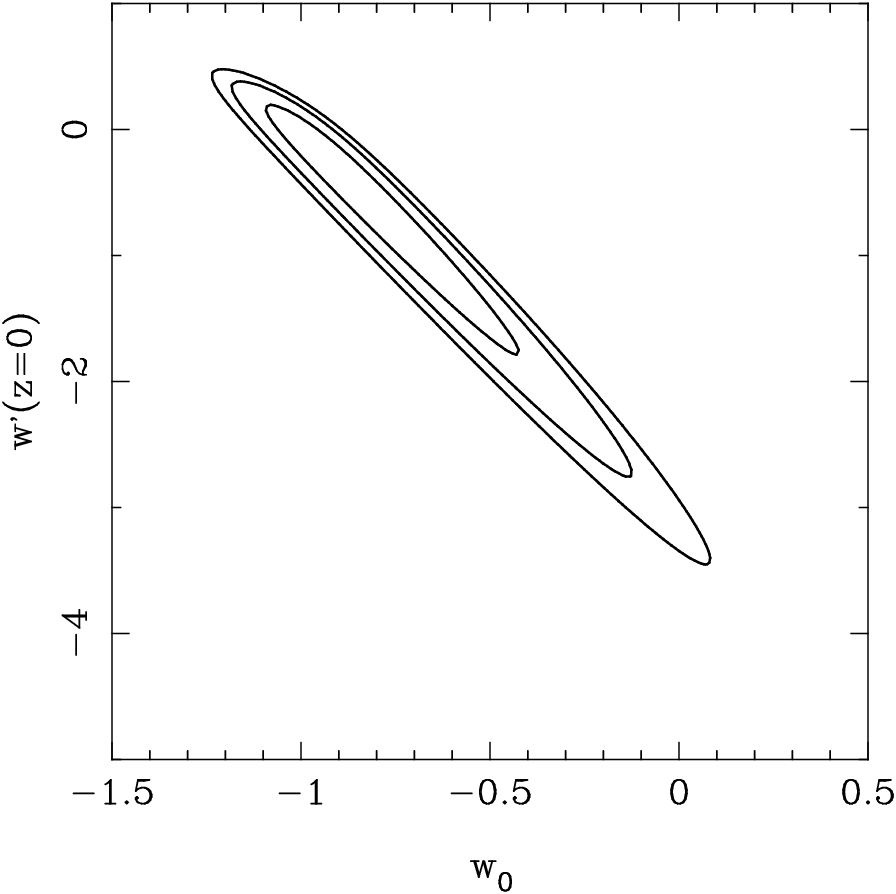}\includegraphics[width=0.2\textwidth]{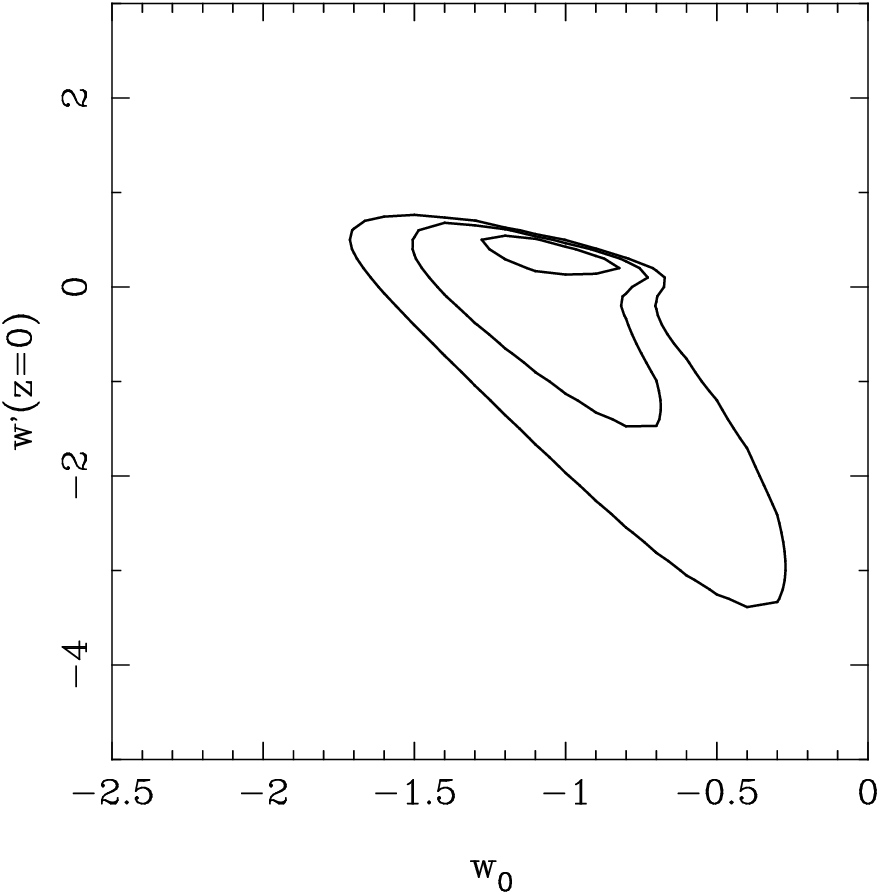}\includegraphics[width=0.2\textwidth]{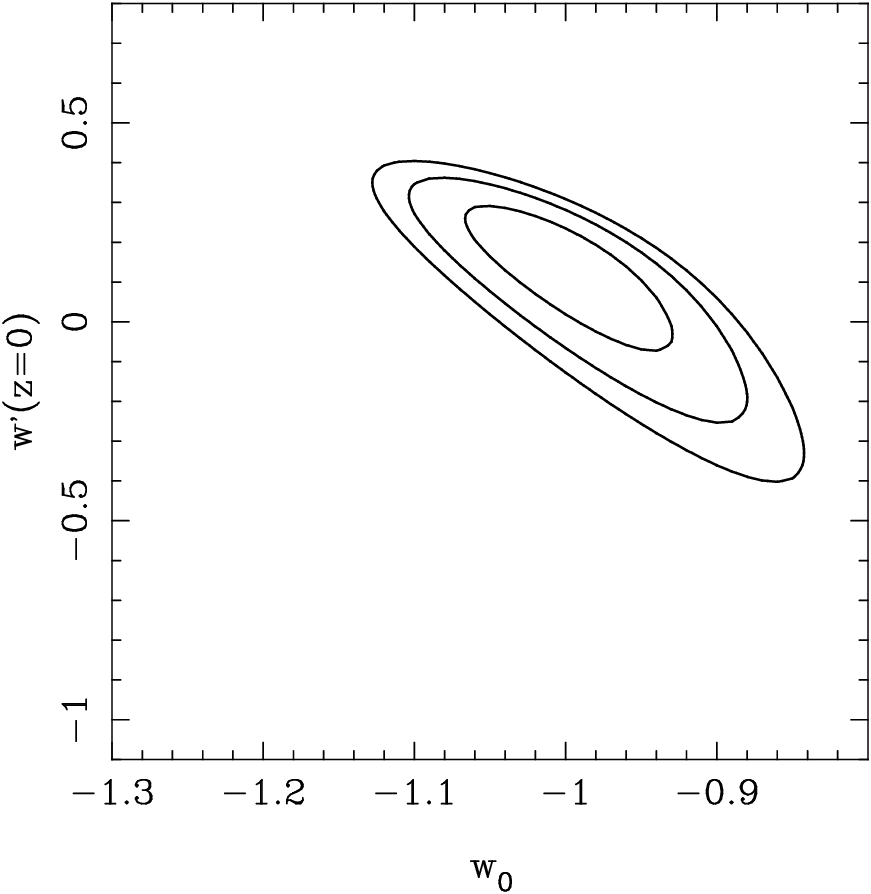}\\

%\end{tabular}
\caption{\label{fig::marginparo}   The plots 
  represent the confidence contours for (from
  left to right) SNIa, BAO, H(z) and a combination of the datasets for
  the three different parameterizations with marginalization over $\Omega_m$. 
  The contours in top row and the second row are  for
  parameterizations $p=1$ and 
  $p=2$ respectively and the last row corresponds to the logarithmic paramterization.
 } 

\end{figure}

\begin{figure}
\centering
%\begin{tabular}{cccc}
\includegraphics[width=0.2\textwidth]{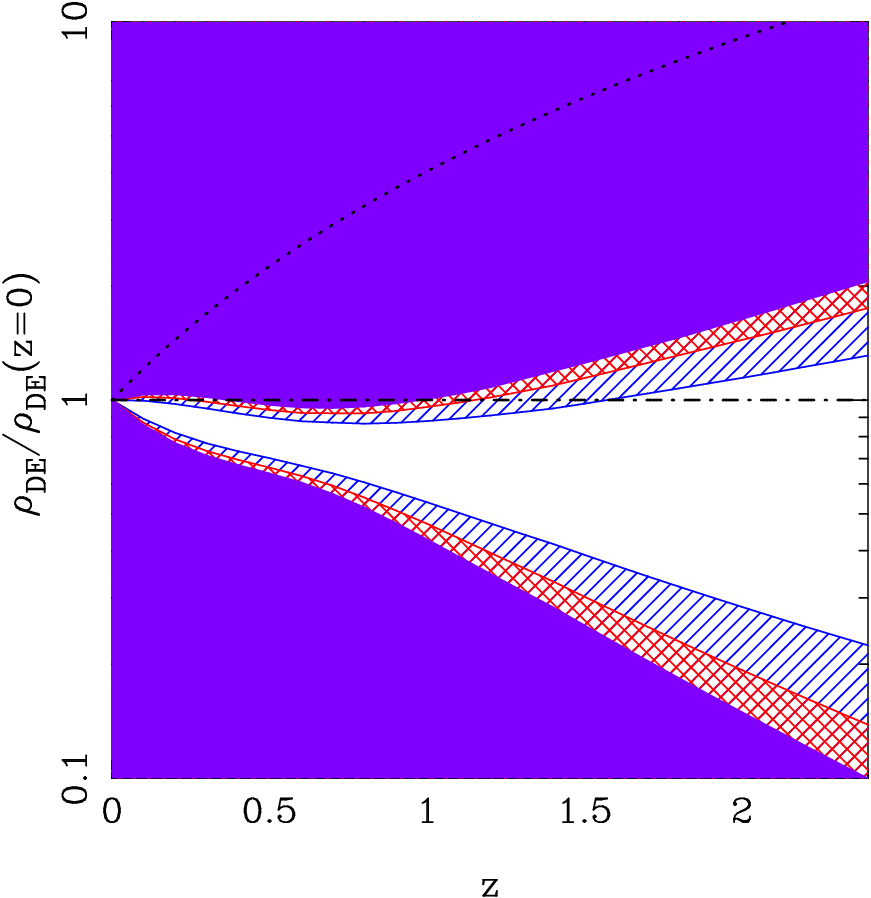}\includegraphics[width=0.2\textwidth]{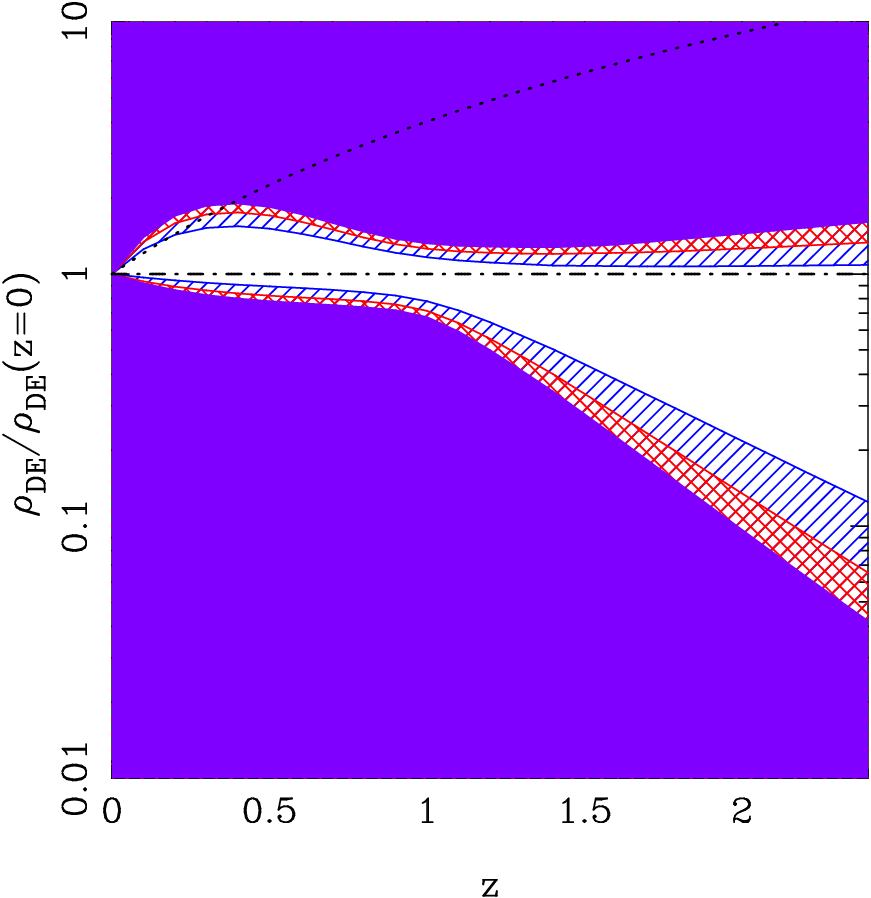}\includegraphics[width=0.2\textwidth]{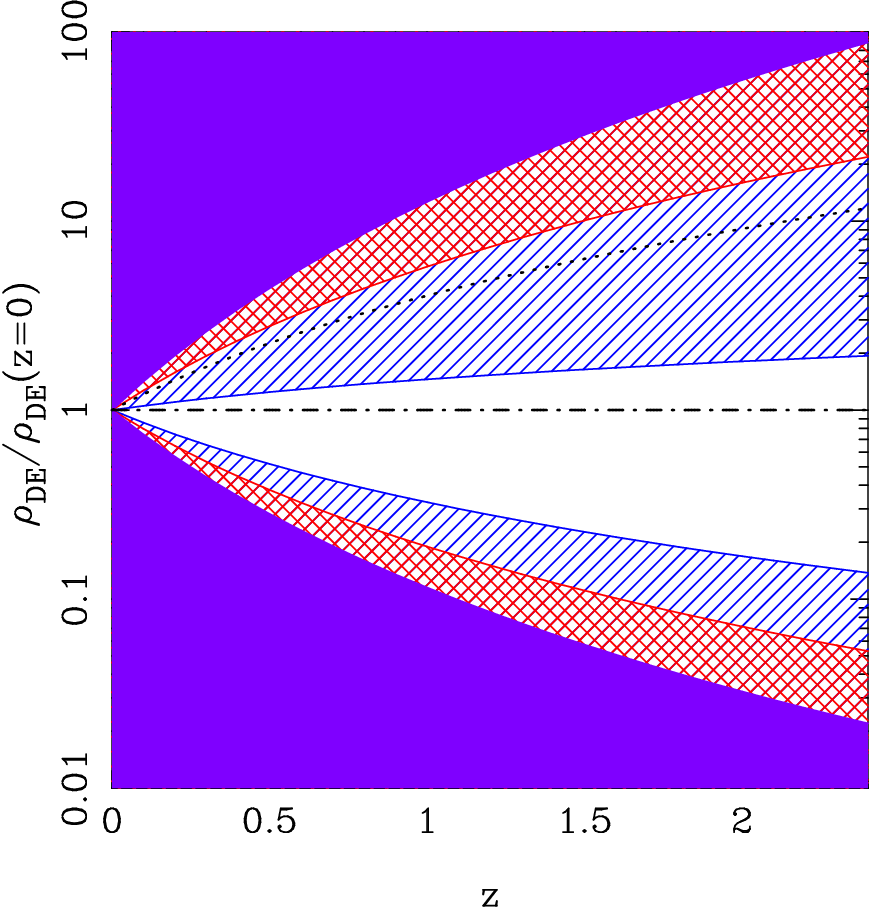}\includegraphics[width=0.2\textwidth]{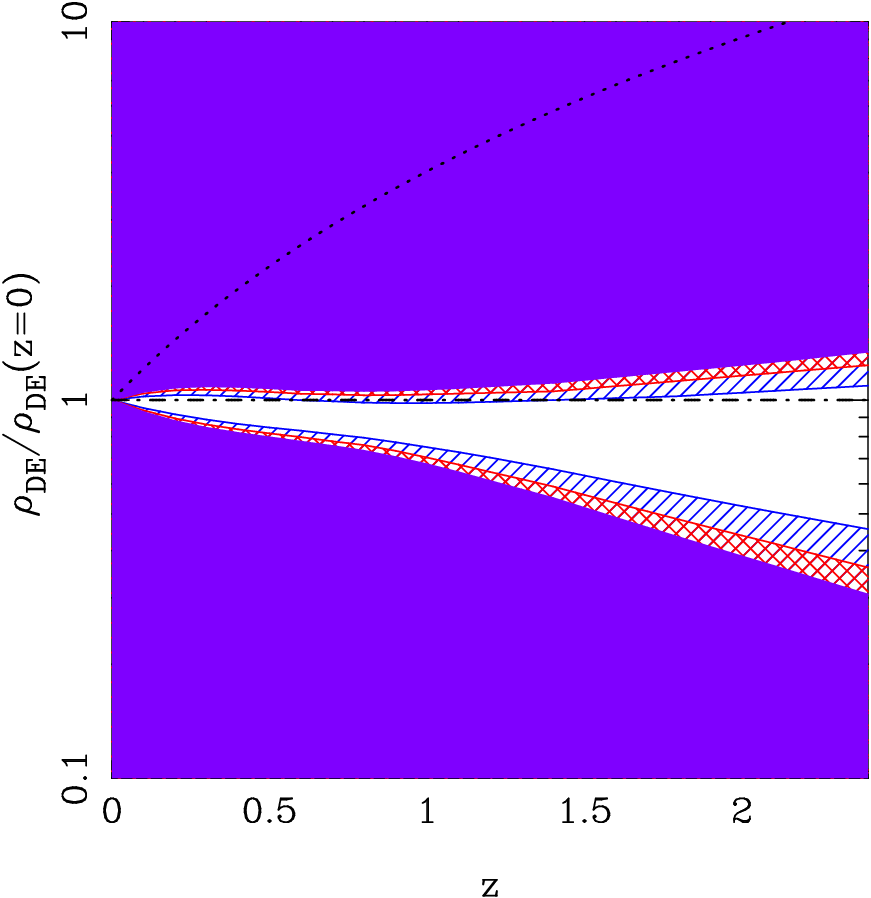}\\
\includegraphics[width=0.2\textwidth]{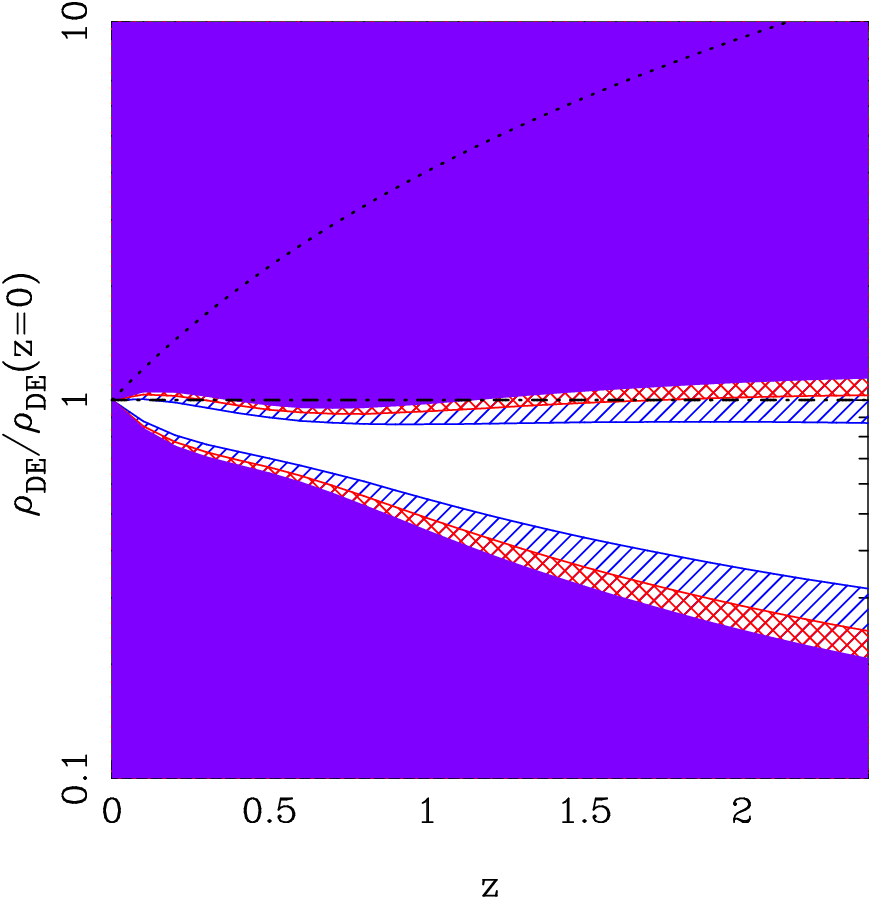}\includegraphics[width=0.2\textwidth]{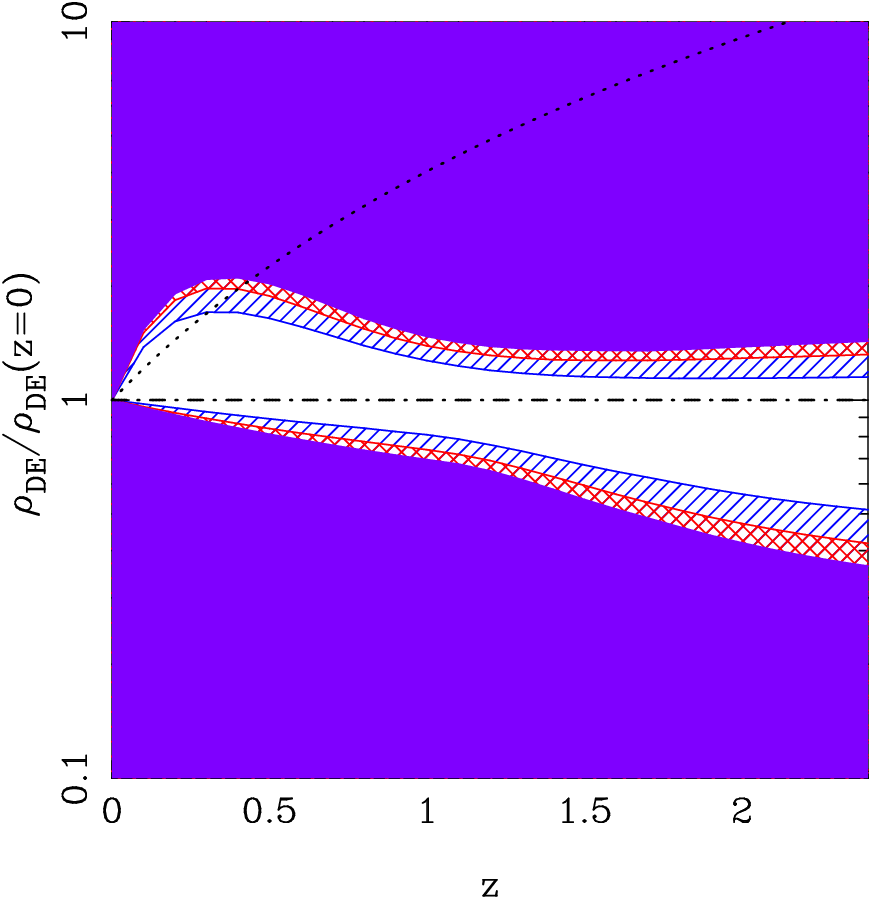}\includegraphics[width=0.2\textwidth]{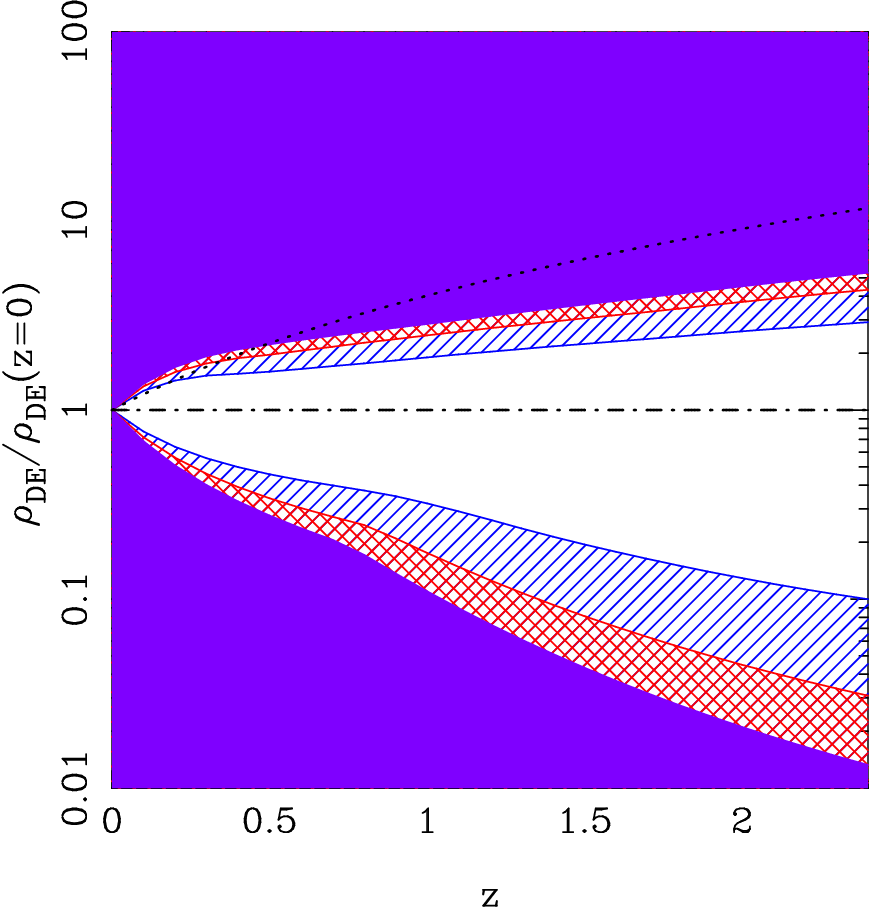}\includegraphics[width=0.2\textwidth]{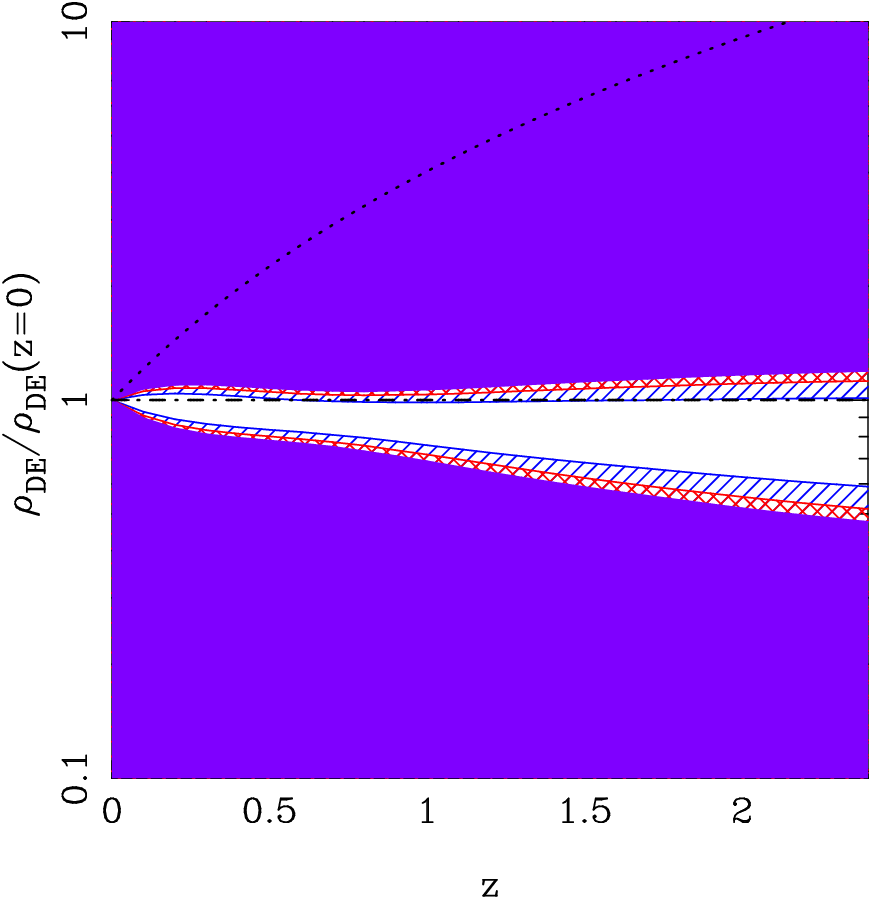}\\
\includegraphics[width=0.2\textwidth]{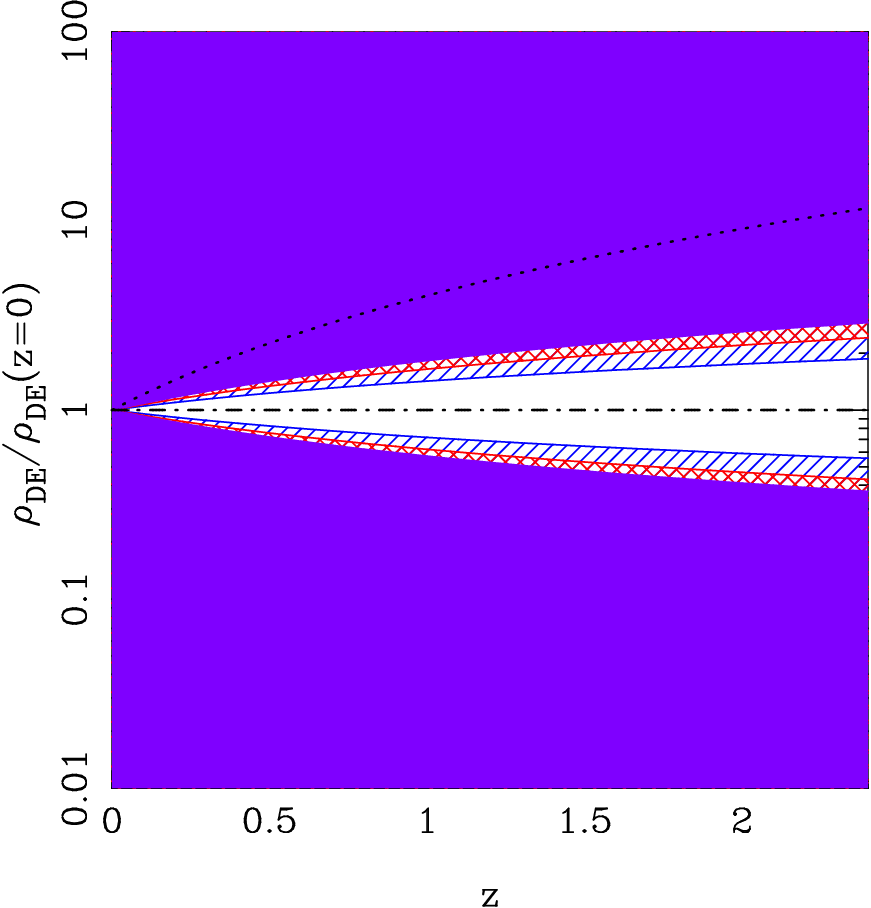}\includegraphics[width=0.2\textwidth]{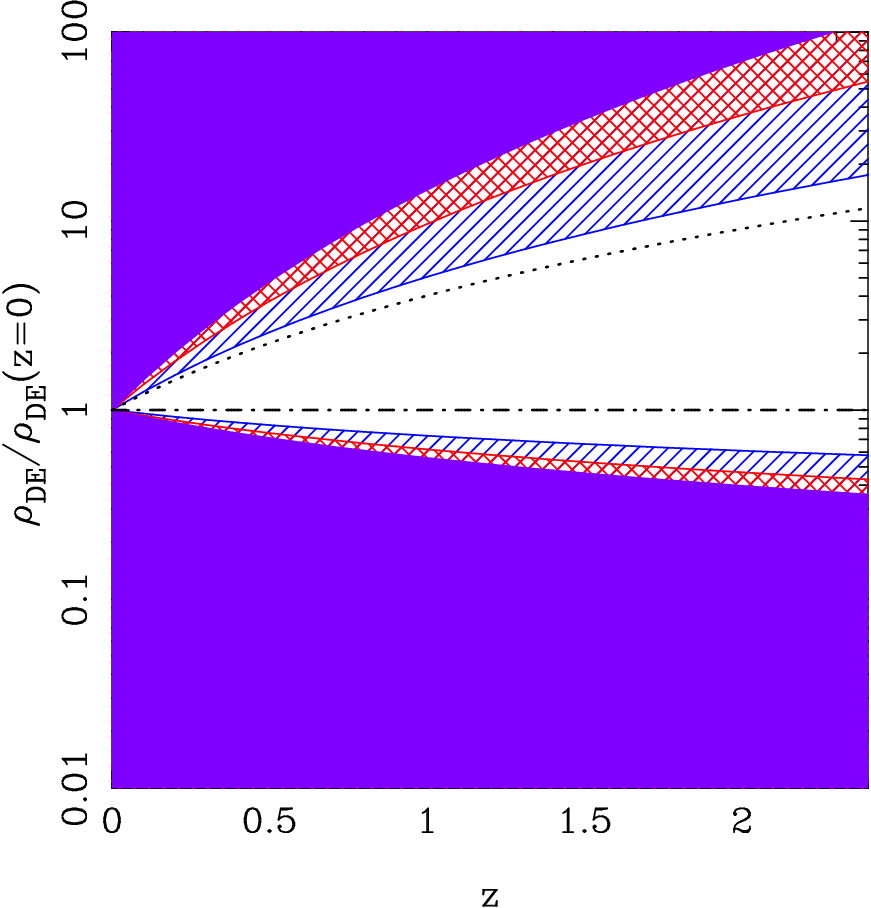}\includegraphics[width=0.2\textwidth]{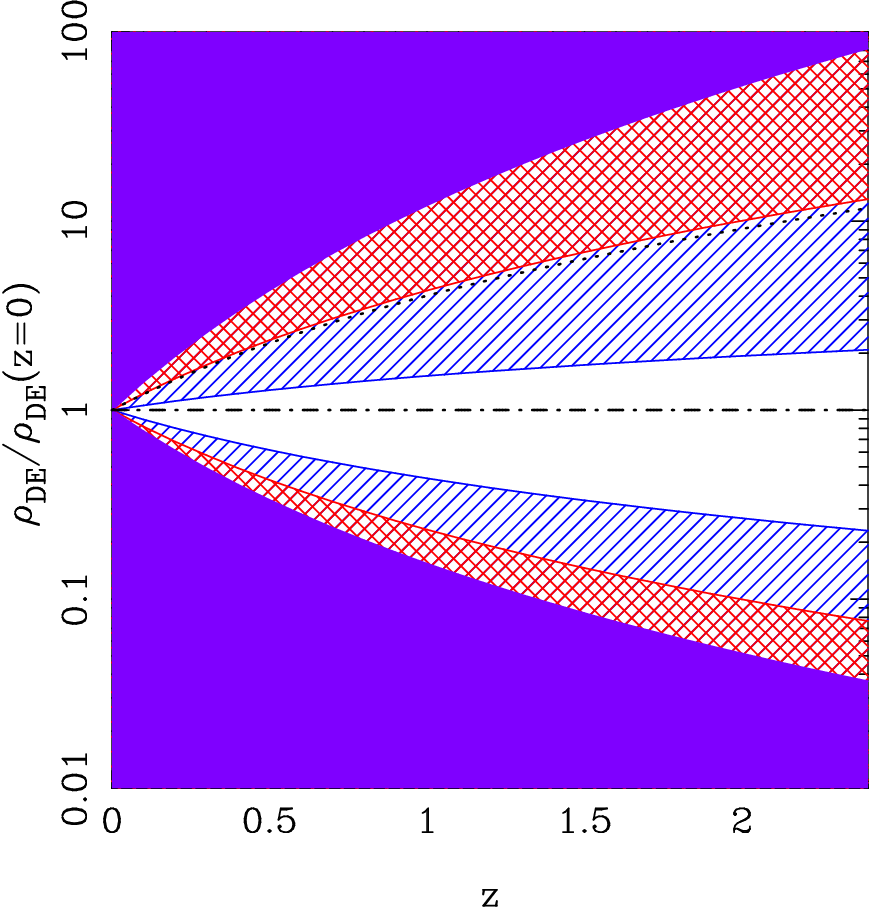}\includegraphics[width=0.2\textwidth]{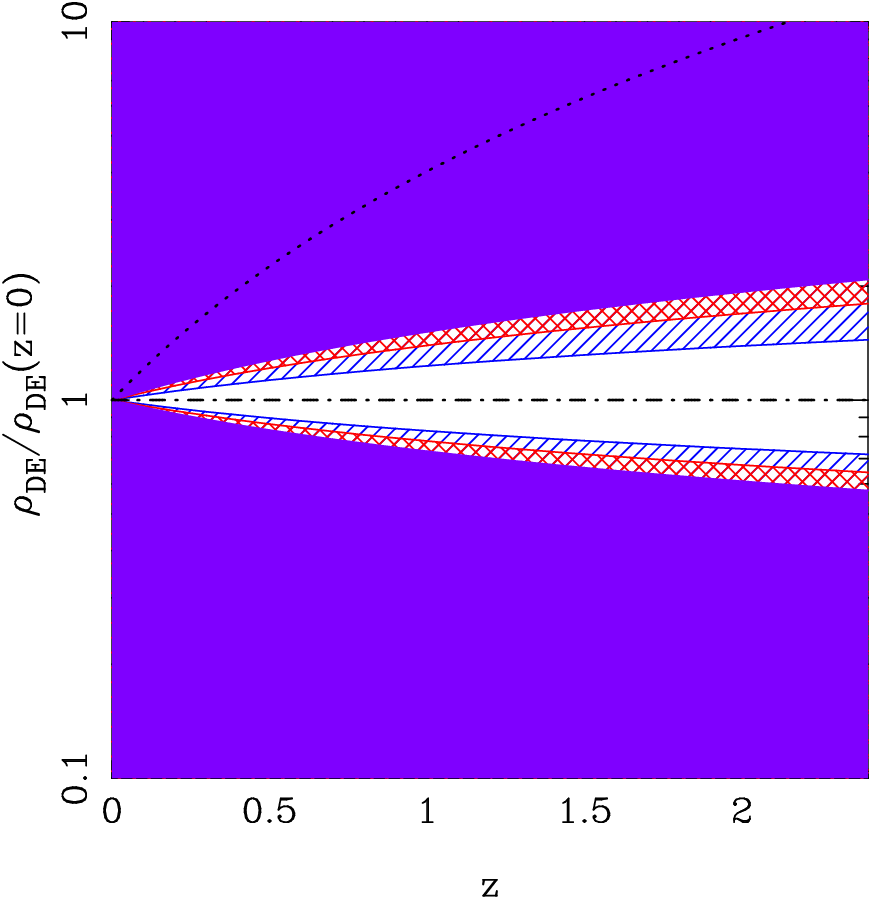}\\

%\end{tabular}
\caption{\label{fig::marginparo1}   The plots show the allowed range of dark energy density by  SNIa, BAO, H(z) and a combined datasets for  
  the three different parameterizations with marginalization over $\Omega_m$.} 
\end{figure}

\begin{figure}
\centering
%\begin{tabular}{cccc}
\includegraphics[width=0.2\textwidth]{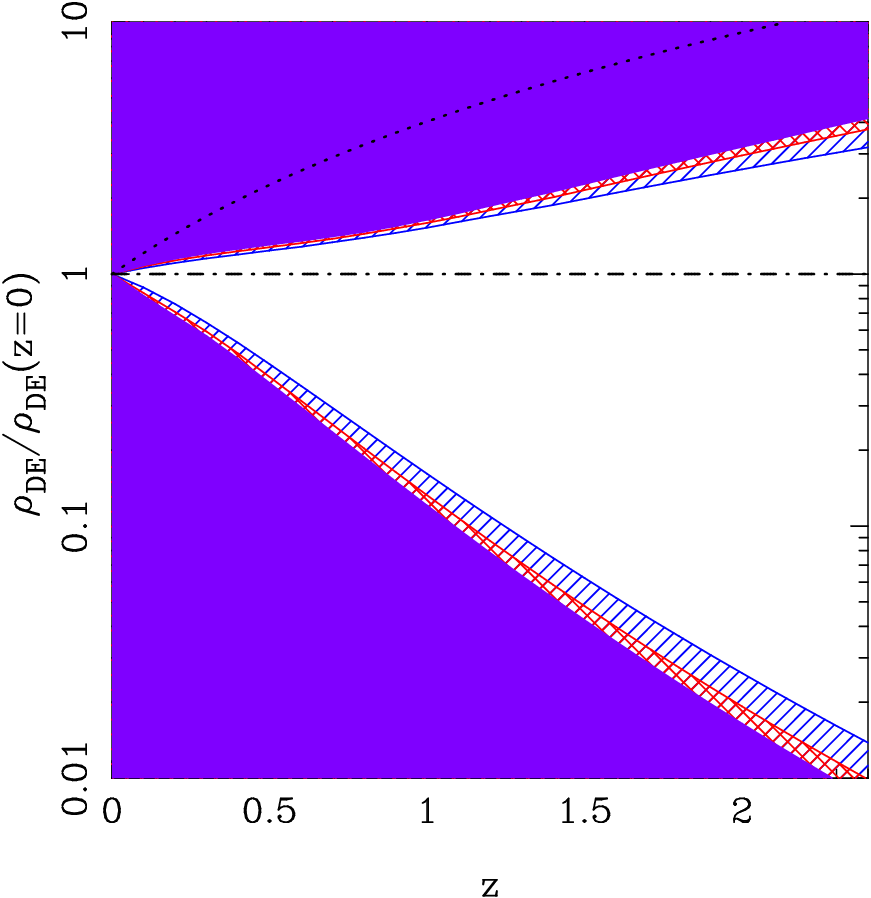} \includegraphics[width=0.2\textwidth]{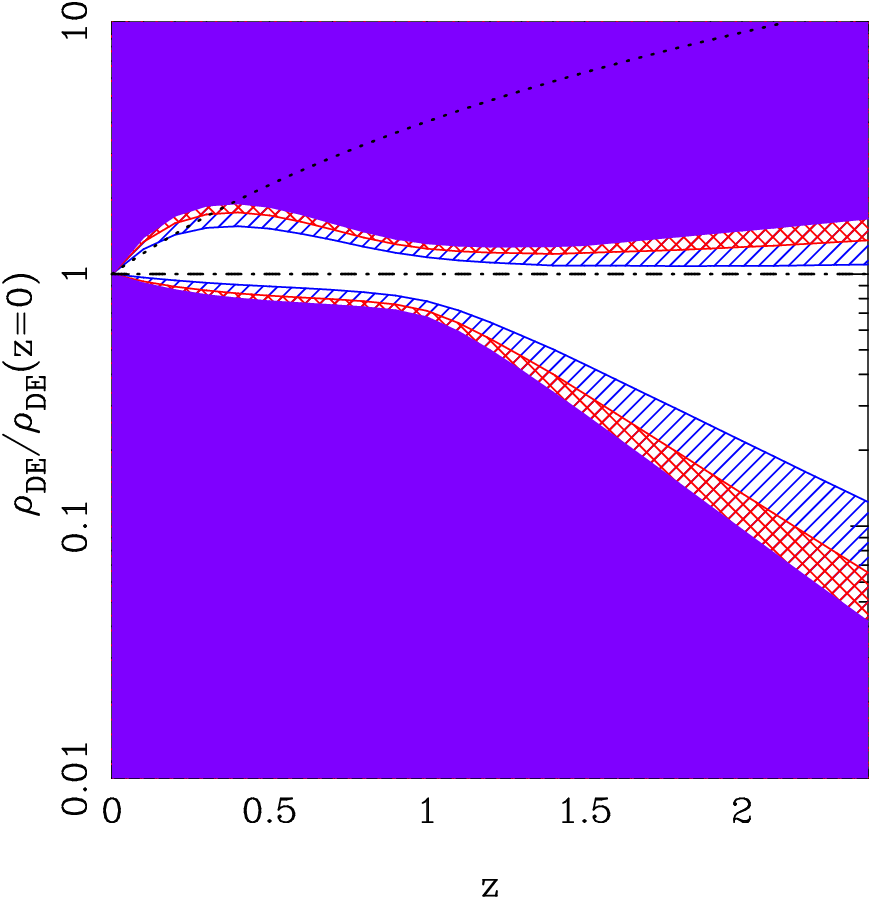} \includegraphics[width=0.2\textwidth]{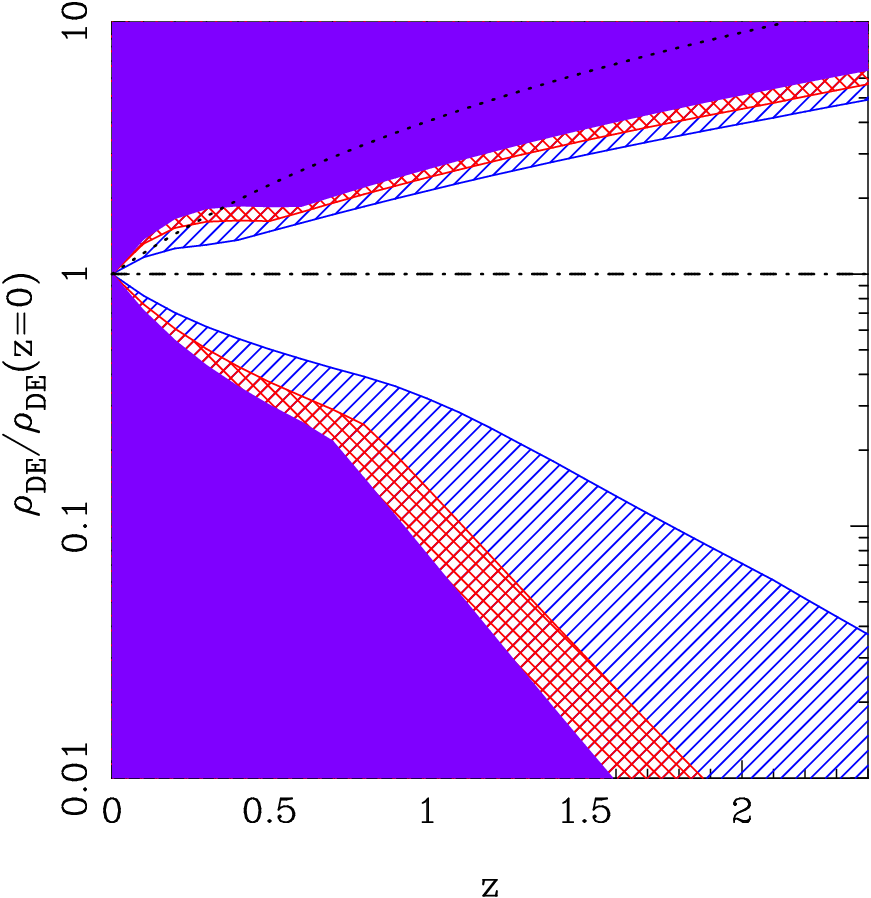} \includegraphics[width=0.2\textwidth]{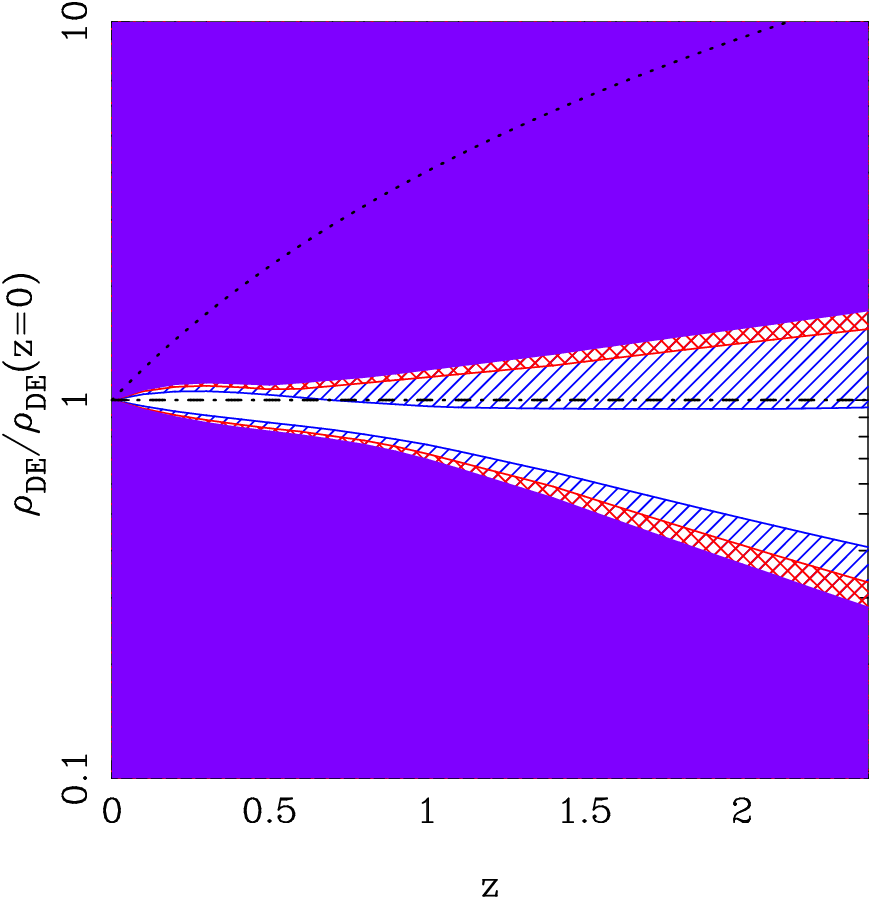}\\

\includegraphics[width=0.2\textwidth]{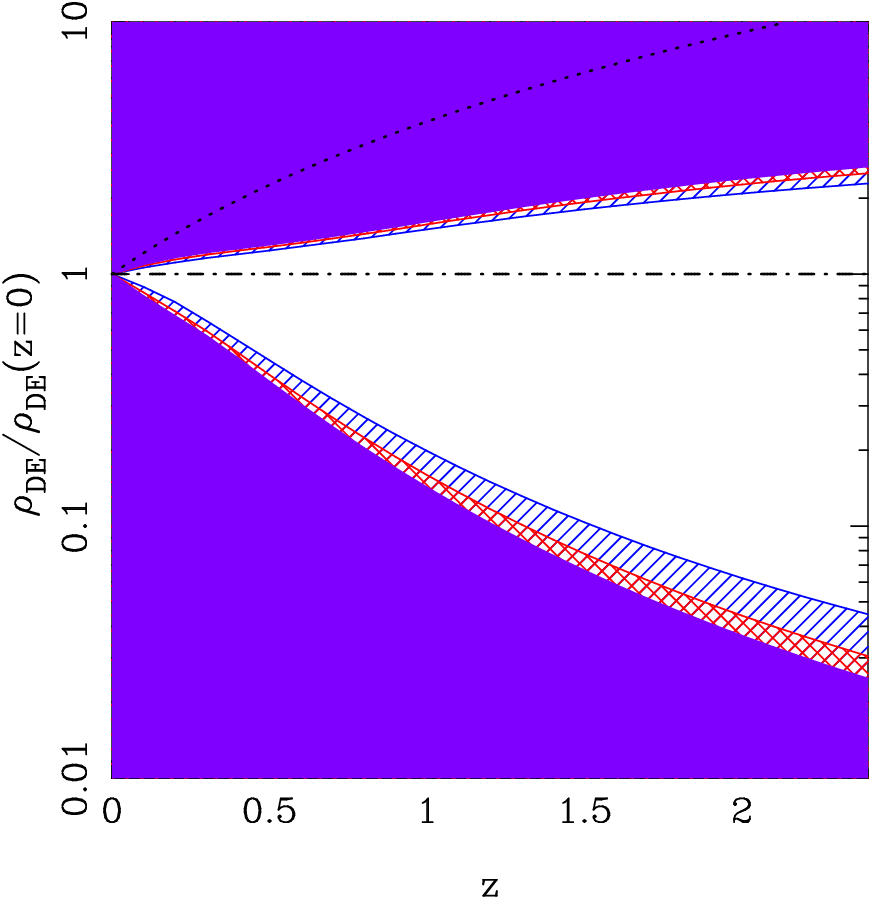} \includegraphics[width=0.2\textwidth]{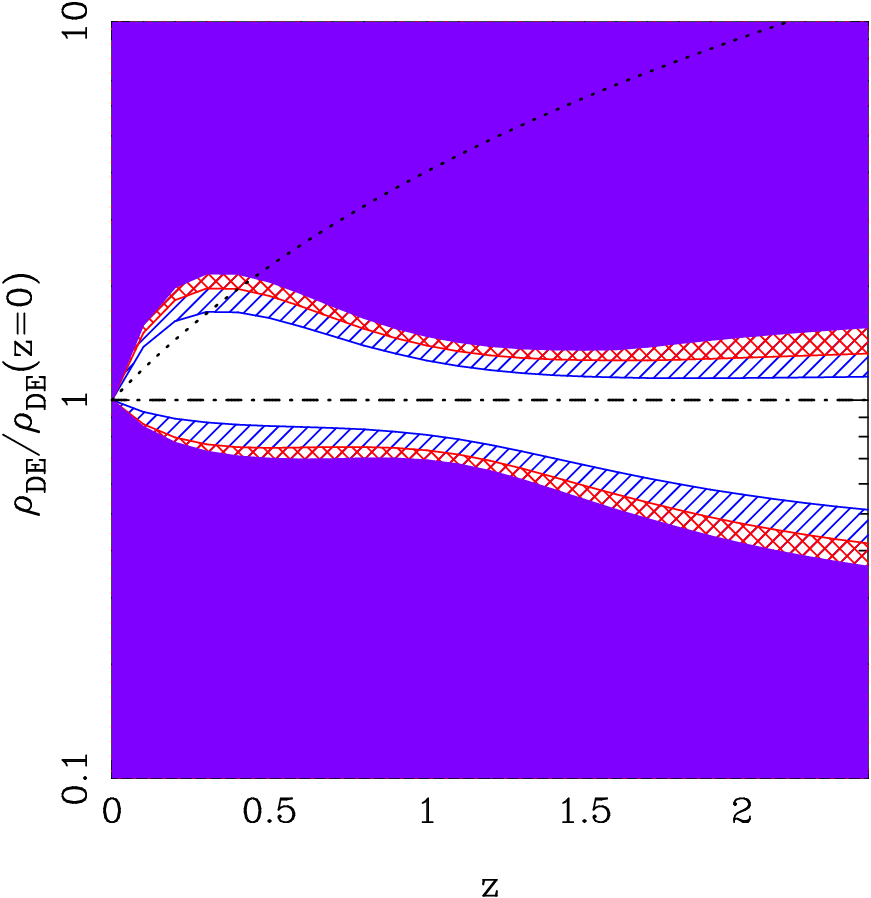} \includegraphics[width=0.2\textwidth]{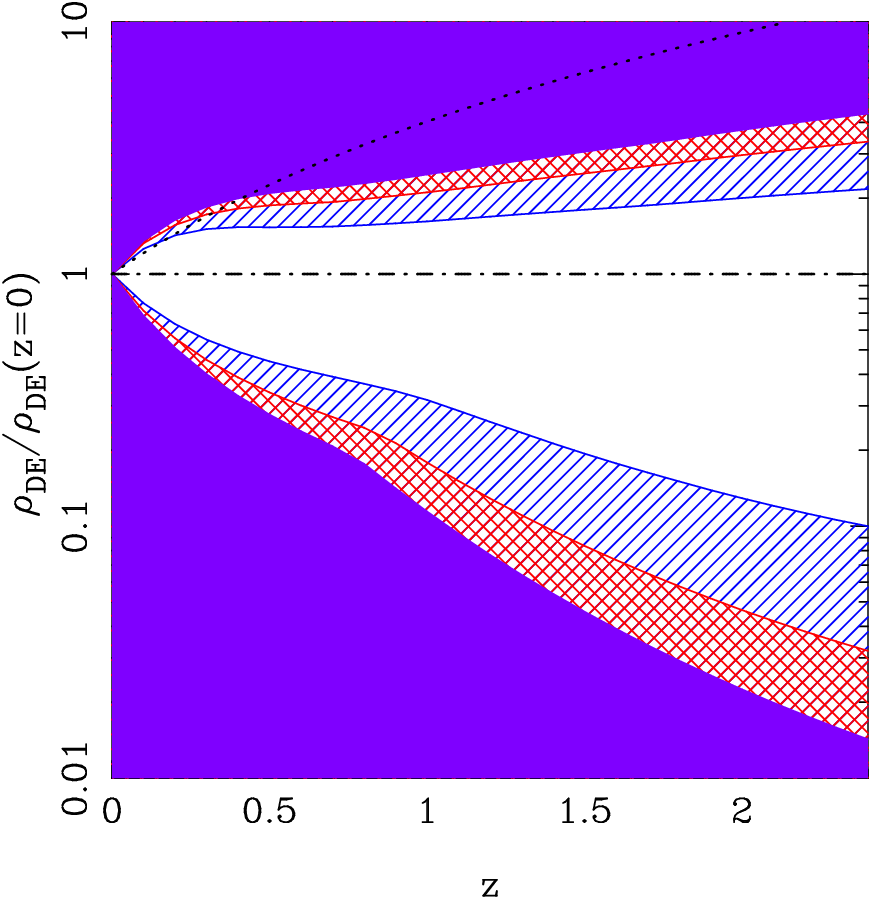} \includegraphics[width=0.2\textwidth]{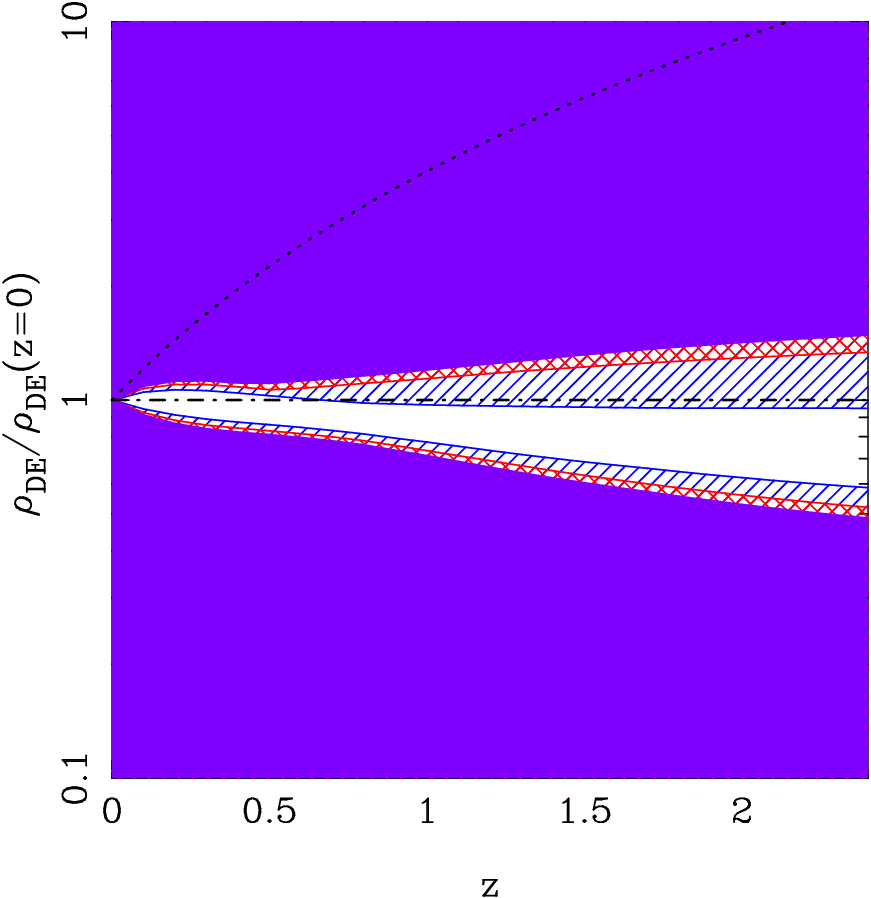}\\

\includegraphics[width=0.2\textwidth]{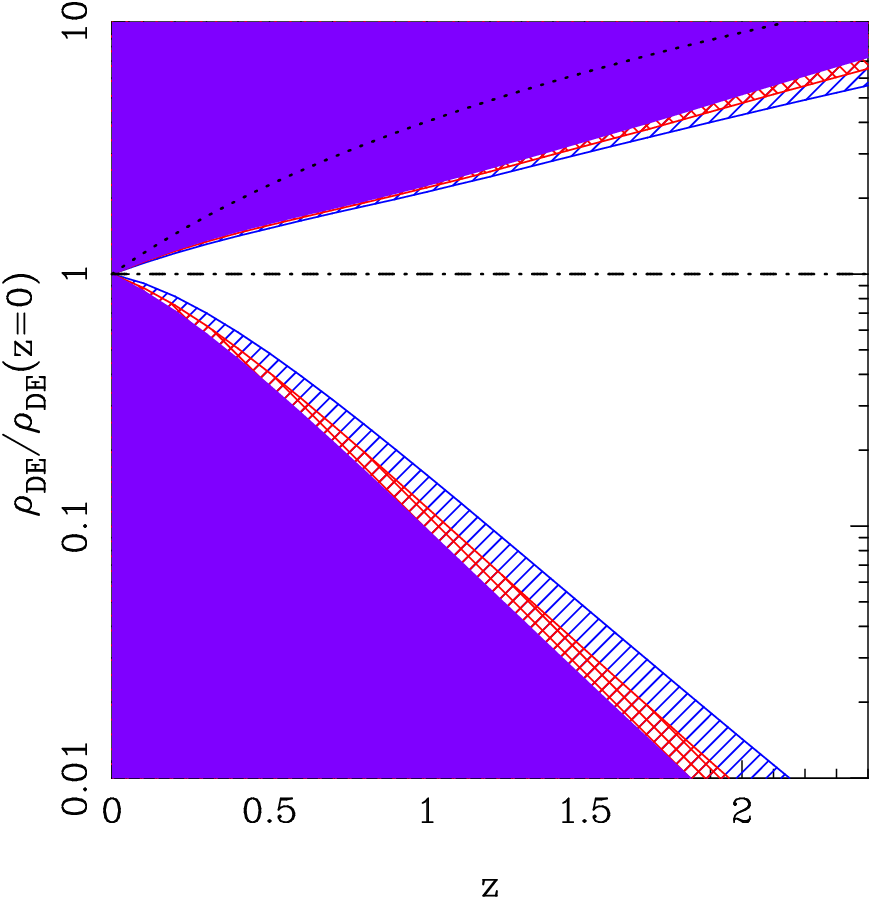} \includegraphics[width=0.2\textwidth]{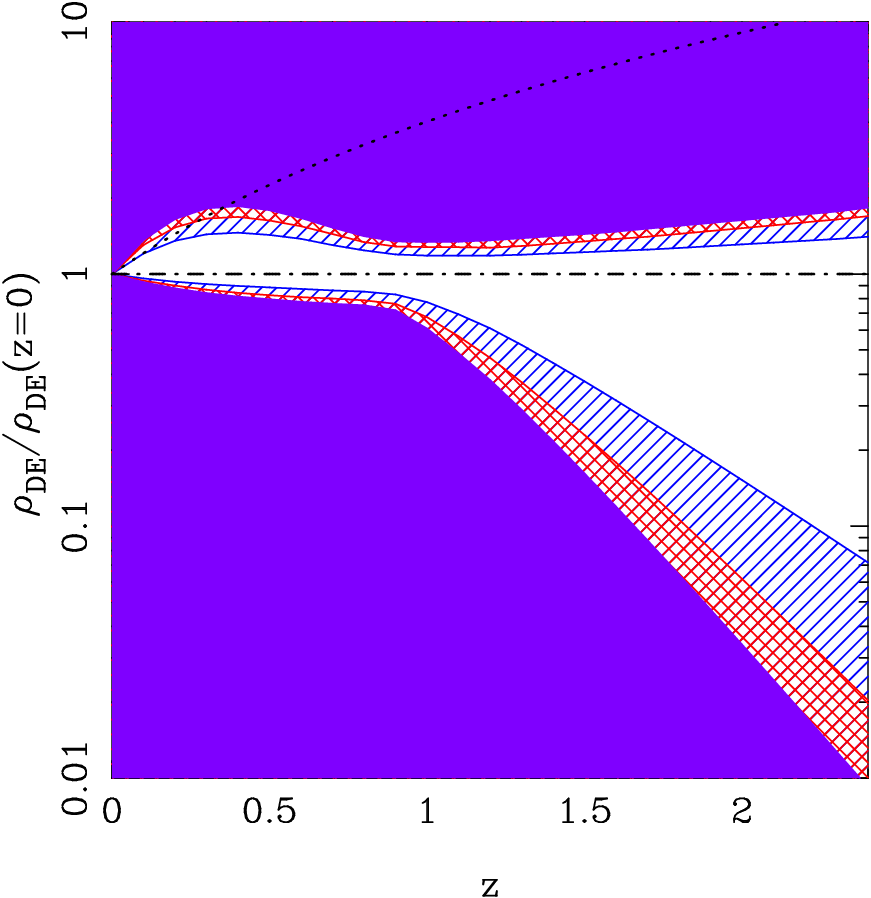} \includegraphics[width=0.2\textwidth]{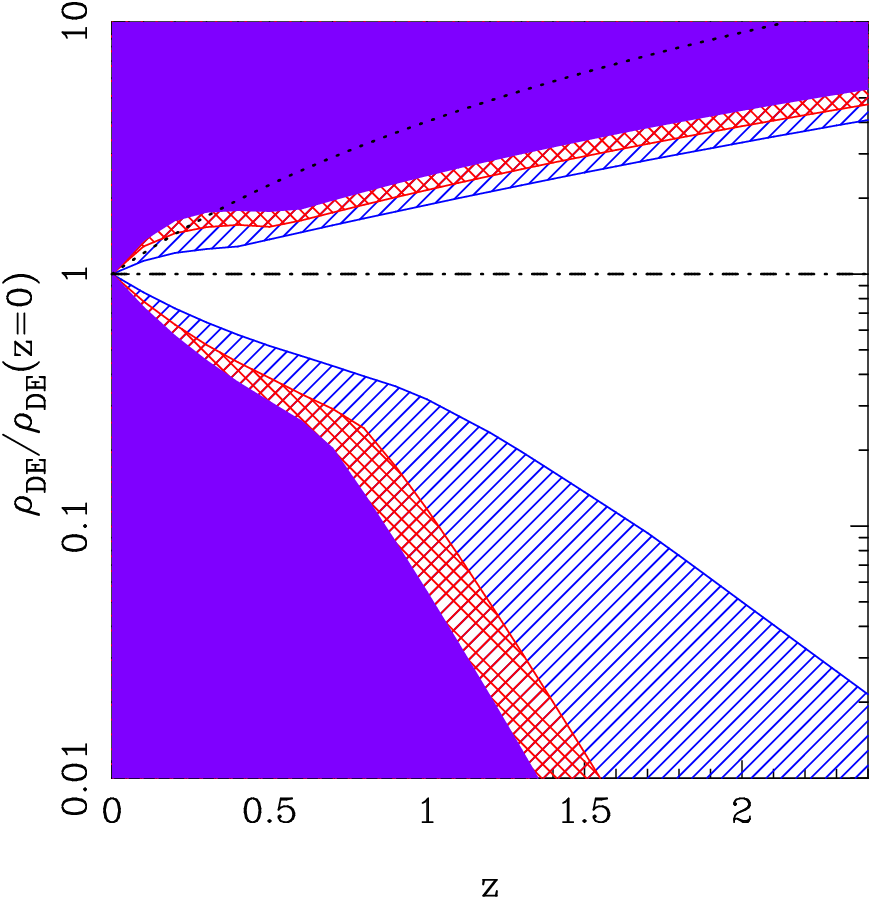} \includegraphics[width=0.2\textwidth]{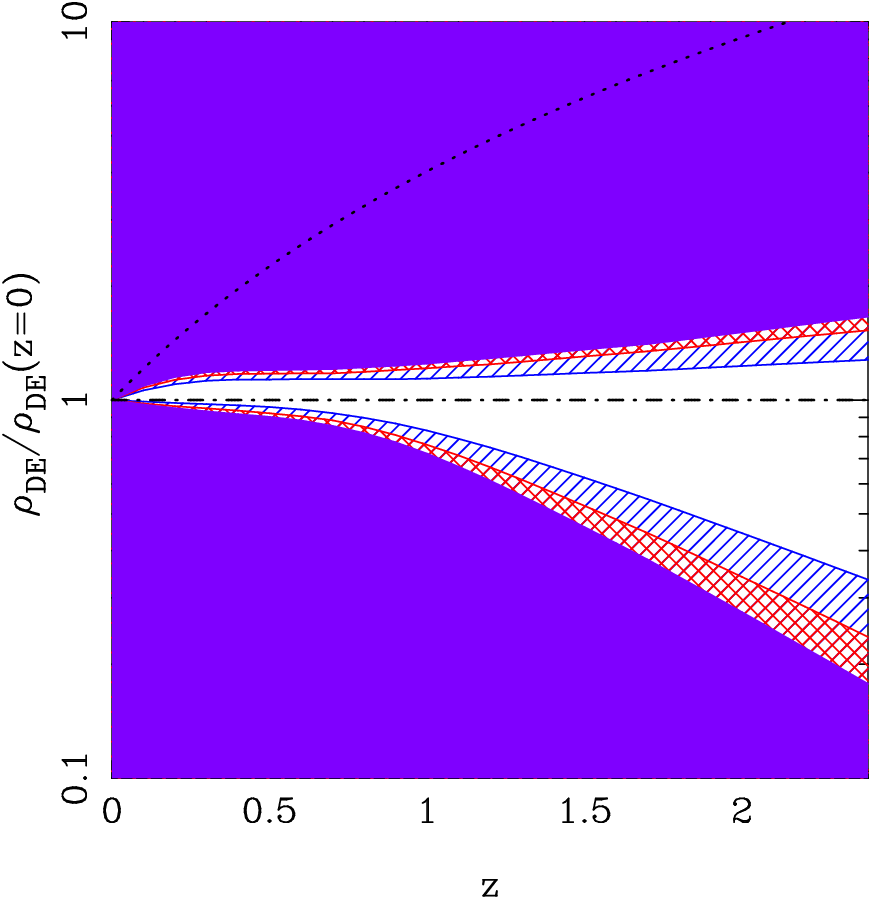}\\

%\end{tabular}
\caption{\label{fig::nomarg}   The plots show the allowed range of dark energy density by  SNIa, BAO, H(z) and a combined datasets for the three different parameterizations with all parameters free.}  

\end{figure}

\begin{table*}
\begin{tabular}{|l|l|l|l|} \hline
Data set&$3\sigma$ confidence&$\chi_m^2$&Best Fit Model \\ \hline
SNIa&0.05$\leq$$\Omega_{m}$$\leq$0.43&562.29&$\Omega_{m}$=0.29\\ 
& -1.57 $\leq$ w $\leq$ -0.66&&w=-1.04\\
&&&\\ \hline
BAO&0.19$\leq$$\Omega_{m}$$\leq$0.36&0.96&$\Omega_{m}$=0.27\\ 
& -2.19 $\leq$ w $\leq$ -0.42&&w=-1.17\\ 
&&&\\ \hline
H(z)&0.2$\leq$$\Omega_{m}$$\leq$0.35&16.27&$\Omega_{m}$=0.28\\ 
& -1.78 $\leq$ w $\leq$ -0.72&&w=-1.12\\
&&&\\ \hline
SNIa+BAO+H(z)&0.25$\leq$$\Omega_{m}$$\leq$0.31&580.81&$\Omega_{m}$=0.28\\ 
& -1.13 $\leq$ w $\leq$ -0.95&&w=-1.03\\
&&&\\ \hline

\end{tabular}
\caption{\label{table::ranges_c}
This table shows the 3$\sigma$ confidence limit for  three
  datasets, for cosmological models where dark energy parameter is a
  constant ($wCDM$  cosmology).}

\end{table*}

\begin{table*}
\begin{tabular}{|l|l|l|l|} \hline
  Data set&$3\sigma$ confidence&$\chi_m^2$&Best Fit Model \\
  \hline
\multicolumn{4}{|c|}{$p=1$} \\
\hline
SNIa & -1.64 $\leq$ $w_0$ $\leq$ -0.72&&$w_0$=-1.0\\ 
& -2.0 $\leq$ $w'(z=0)$ $\leq$ 1.26&562.25&$w'(z=0)$=0.2\\
& 0.2 $\leq$ $\Omega_m$ $\leq$ 0.45&&$\Omega_m$=0.25\\
 \hline
BAO& -1.3 $\leq$ $w_0$ $\leq$ 0.33&&$w_0$=-0.67\\ 
& -4.97 $\leq$ $w'(z=0)$ $\leq$ 0.77&2.13&$w'(z=0)$=-1.26\\
& 0.3 $\leq$ $\Omega_m$ $\leq$ 0.31&&$\Omega_m$=0.3\\
\hline
H(z)& -2.14 $\leq$ $w_0$ $\leq$ 0.28&&$w_0$=-1.16\\ 
& -5.0 $\leq$ $w'(z=0)$ $\leq$ 1.8&20.77&$w'(z=0)$=0.8\\
& 0.1 $\leq$ $\Omega_m$ $\leq$ 0.37&&$\Omega_m$=0.2\\
 \hline
SNIa+BAO+H(z)& -1.2 $\leq$ $w_0$ $\leq$ -0.74&&$w_0$=-1.0\\ 
& -1.32 $\leq$ $w'(z=0)$ $\leq$ 0.56&585.67&$w'(z=0)$=-0.26\\
& 0.25 $\leq$ $\Omega_m$ $\leq$ 0.3&&$\Omega_m$=0.3\\
 \hline
\multicolumn{4}{|c|}{$p=2$} \\
\hline
SNIa& -1.62 $\leq$ $w_0$ $\leq$ -0.62&&$w_0$=-1.06\\ 
& -3.0 $\leq$ $w'(z=0)$ $\leq$ 2.56&562.28&$w'(z=0)$=-0.06\\
& 0.2 $\leq$ $\Omega_m$ $\leq$ 0.45&&$\Omega_m$=0.3\\ 
 \hline
BAO& -1.68 $\leq$ $w_0$ $\leq$ 0.99&&$w_0$=-0.44\\ 
& -10.0 $\leq$ $w'(z=0)$ $\leq$ 2.99&2.79&$w'(z=0)$=-3.06\\
& 0.25 $\leq$ $\Omega_m$ $\leq$ 0.3&&$\Omega_m$=0.3\\
 \hline
H(z)& -2.36 $\leq$ $w_0$ $\leq$ 0.24&&$w_0$=-1.16\\ 
& -5.8 $\leq$ $w'(z=0)$ $\leq$ 3.4&21.11&$w'(z=0)$=0.3\\
& 0.18 $\leq$ $\Omega_m$ $\leq$ 0.37&&$\Omega_m$=0.28\\
 \hline
SNIa+BAO+H(z)&-1.36 $\leq$ $w_0$ $\leq$ -0.64&&$w_0$=-1.0\\ 
& -2.52 $\leq$ $w'(z=0)$ $\leq$ 1.62&586.46&$w'(z=0)$=-0.38\\
& 0.25 $\leq$ $\Omega_m$ $\leq$ 0.3&&$\Omega_m$=0.3\\
 \hline
\multicolumn{4}{|c|}{Logarithmic} \\
\hline
SNIa& -1.44 $\leq$ $w_0$ $\leq$ -0.58&&$w_0$=-0.94\\ 
& -2.0 $\leq$ $w'(z=0)$ $\leq$ 0.68&563.02&$w'(z=0)$=-1.04\\
& 0.1 $\leq$ $\Omega_m$ $\leq$ 0.49&&$\Omega_m$=0.37\\ 
 \hline
BAO& -1.26 $\leq$ $w_0$ $\leq$ 0.2&&$w_0$=-0.84\\ 
& -3.8 $\leq$ $w'(z=0)$ $\leq$ 0.5&1.29&$w'(z=0)$=-0.5\\
& 0.26 $\leq$ $\Omega_m$ $\leq$ 0.32&&$\Omega_m$=0.29\\
 \hline
H(z)& -2.0 $\leq$ $w_0$ $\leq$ 0.2&&$w_0$=-1.1\\ 
& -5.0 $\leq$ $w'(z=0)$ $\leq$ 0.9&20.91&$w'(z=0)$=0.3\\
& 0.1 $\leq$ $\Omega_m$ $\leq$ 0.37&&$\Omega_m$=0.24\\
 \hline
SNIa+BAO+H(z)&-1.09 $\leq$ $w_0$ $\leq$ -0.66&&$w_0$=-0.91\\ 
& -1.21 $\leq$ $w'(z=0)$ $\leq$ 0.25&587.18&$w'(z=0)$=-0.29\\
& 0.26 $\leq$ $\Omega_m$ $\leq$ 0.32&&$\Omega_m$=0.29\\
 \hline
\end{tabular}
\caption{\label{table::ranges_no_marg}This table shows the 3$\sigma$ confidence limit for various data sets for the different varying dark energy models for $H_0=70$ $km/s/Mpc$ for BAO and SNIa and for H(z) data, marginalized over $H_0$. For these results, we have  kept $\Omega_m$ a free parameter.}

\end{table*}

\section{Analysis and Results} \label{sec::results}

The SNIa UNION compilation data \cite{sn9}
comprises of  distance moduli of  $580$ supernovae upto 
$z \sim 1.4$. The BAO observations are compiled from a variety of
observations \cite{bao00,bao1, bao2, bao3,bao4,new_bao1, new_bao2} and cover redshift up to $z=2.3$.   
The third data we use are estimates of the Hubble parameter H(z) at
different redshifts \cite{hz1,hz2,hz3,hz4,hz5,hz6,hz7,hz8,hz9} in the range $0.07\leq z\leq
2.3$. 

We use the $\chi^{2}$ minimization technique to analyze the data and
constrain cosmological parameters of interest in this work.  
For each data set, the observed quantity  $X_{i,ob}$ at a certain redshift
$z_i$, is compared with that obtained theoretically, $X_{i,th}$,
at the same redshift, for each class of dark energy models.
The $\chi^2$ is  defined as
\begin{eqnarray}\label{chisq}
\chi^2 = \sum_i\left[\frac{X_{i,ob}-X_{i,th}}{\sigma_i}\right]^2
\end{eqnarray}

The absolute magnitude of supernova is a nuisance parameter as far as
dark energy parameters are concerned, and as the determination of
absolute determination is degenerate with determination of value of
Hubble constant, and the latter is also then a nuisance parameter.
We consider the distance modulus given in \cite{sn9} and
for consistency with the earlier analyses we fix the value of $H_0$ to
$70~km/s/Mpc$ (see also, \cite{hz10}).
Similarly, while analyzing the BAO observations, the $D_v(z)$ and
$A(z)$ have been determined using $H_0$= 70 $km/s/Mpc$.
We are interested primarily in dark energy parameters, hence we
marginalize over the present value of the Hubble parameter while
analyzing $H(z)$ \cite{hz1,hz2,hz3,hz4,hz5,hz6,hz7,hz8,hz9}.
We consider the range $H_0=65-75~km/s/Mpc$ for this analysis and
marginalize over this range as mentioned above.
The priors are based on the results obtained using earlier,
independent observations.
For $w$CDM model, the priors used are $\Omega_{m}$=$0.01-0.6$, and
$w=-4.0- 0$.
The priors for this analysis are listed in Table.~\ref{table::priors}. 
For models with a varying equation of state parameter, the present day
value of dark energy equation of state $w_0$ is taken in the range
$-5.0$ to $2.0$  and $w'(z=0)$ varies from $-10.0$ to $10.0$.
These priors are listed  in Table.~\ref{table::priors_two}.

In Fig.~\ref{fig::constw},  we show the constraints using these datasets 
for a cosmological model with constant equation of state parameter for
dark energy ($w$CDM model).  
This figure shows $1 \sigma$, $2 \sigma$ and $3 \sigma$ contours
corresponding to $67.3 \%$, $95 \%$ and $99 \%$ confidence levels in
the $\Omega_{m}$-$w$ plane.
Going from left to right, the plots correspond to  the confidence
contours obtained by analyzing SNIa data, BAO data, H(z) data and
their combination, respectively.
The minimum value $\chi_{m}^2= 562.29$ corresponds to the values $w =
-1.04$  and $\Omega_{m} = 0.29$ for SNIa data.
In case of the BAO dataset, $\chi_{m}^2=0.96$ for $w = -1.2$ and
$\Omega_{m} = 0.27$.
With $H(z)$ data, $\chi_{m}^2=16.27$ for $w = -1.12$ and $\Omega_{m} = 0.28$. 
It is apparent from the figure that the BAO observations provide
constraints which are complementary to those obtained from the other
two observations and hence when combined, the allowed range of
cosmological parameters is constrained very strongly. 
In this set of models with a constant equation of state parameter, the
combined constraints are consistent with a cosmological constant. 
The confidence limits for these results are listed in Table
~\ref{table::ranges_c}.   

In dark energy models where $w$ is a function of redshift, there are
four parameters: $\Omega_{m}$, $w_0$, $w'(z=0)$ and $H_0$.
The priors for these parameters are listed in Table \ref{table::priors_two}.
In Figs. \ref{fig::marginpar} and \ref{fig::marginpar1}, we show
confidence contours in  
$\Omega_{m}-w_0$ plane and $\rho_{DE}/\rho_{DE}(z=0)$ respectively with 
marginalization over $w'$.
For H(z) data we have marginalized over $H_0$ and in case of SNIa and
BAO data, $H_0$= $70~km/s/Mpc$.
The dotted line and dashed line (plotted here for reference) in the
energy density plots corresponds to w=$-1/3$ and w=$-1$ respectively.
In Figs. \ref{fig::marginparo} and \ref{fig::marginparo1}, we show
confidence contours in $w'-w_0$ plane and  variation in dark energy density
corresponding to the allowed range in $w_0$ and $w'$  with
marginalization over $\Omega_m$. 

While the allowed ranges in $w_0$ and $w'$ are different for these 
scenarios, the allowed ranges of dark energy density are comparable at
lower redshifts (z$<$1). 
For higher  redshifts(z$>$1), the $p=2$ parameterization allows for a smaller
range of allowed dark energy density.
This is due to the fact that the asymptotic value at large redshifts
approaches $w_0 + w'$ for $p=1$ parameterization 
and for $p=2$ the equation of state parameter reverts to the value
$w_0$ 
A larger range is allowed in the case of the logarithmic
parameterization as the equation of state parameter increases monotonically.  
Since dark energy is subdominant at redshifts larger than unity, this
variation does not make a  significant difference to the observational
constraints.

The SNIa data shows a significant preference for 'phantom'  
models and rules out cosmological constant. The SNIa data prefers higher 
values of $\Omega_{m}$ for both the parameterizations.
The other two datasets are at tension with the supernova data, the BAO
data and the $H(z)$ data prefer 'quintessence'  models and also allow
a non-accelerating universe.
The last two  datasets are consistent with a cosmological model for the two
different parameterizations considered here.
These full range of all the free parameters used in the analysis are 
summarized in Table \ref{table::ranges_no_marg}.
In Fig. \ref{fig::nomarg} we show the allowed range of dark energy
density for this case. 
The allowed variation in dark energy density is larger  to that
obtained by marginalization over $\Omega_m$, however the combined
range is similar in the two cases.
While the individual observations allow a cosmological constant model,
the preference for an equation of state with $w<-1$ by the supernova data
dominates the combined results and disallow a cosmological constant.

\section{Summary and Conclusions} \label{sec::conclusions}

In this paper, we have investigated constraints from  current
observations on dark energy equation of state parameter and its evolution
at low redshifts.
In order to do this analysis, we  considered the Supernova Type Ia
data, Baryon Acoustic Oscillation  data and direct  measurements of
the Hubble parameter H(z) at different redshifts.
We analyse the data separately and also combine the constraints obtained
from these observations.

While  the observations support the concordant cosmological constant model, 
the data does not rule out a constant dark energy which is not a
cosmological constant and also allows some variation in the equation
of state parameter. 
These datasets put comparable limits on evolution of dark energy
density as a function of redshift and combining the datasets leads to
significant reduction in the allowed  range of dark energy density
evolution.
The constraints from the BAO data are complementary to
those obtained by the SNIa data and therefore the combination is more
effective in ruling out a large range of paramters.

We extend the analysis to models with a varying equation of state
parameter and we consider three different functions of redshift.
While the  parameterizations considered here may be
different in their nature, the allowed evolution of dark energy
density is fairly model independent, especially if the observations
are combined.
For those models in which the equation of state parameter is a
function of redshift, there is a tension between the SNIa
data and other data 
sets considered in this paper, as SNIa dataset prefer phantom like
models, with $w<-1$ over the quintessence ($w>-1$).

The Planck best-fit base $\Lambda$CDM cosmology is in
good agreement with results from BAO surveys, and with the recent
Joint Light Curve Analysis (JLA) sample of SNIa. 
The Hubble parameter observations too, in combination with either of these
observations, significantly reduce the allowed range.
The consistency of combined observational constraints with a
cosmological constant is primarily due to the BAO and H(z)
measurement data.  
This is due to the large range of  redshifts spanned by these datasets.
More BAO and $H(z)$ observations at high and
intermediate redshifts would further limit the number of viable
cosmological models in addition to currently available data.

\acknowledgments
HKJ and AT thank Department of Science and Technology (DST), Delhi for funding via 
project SR/FTP/PS-127/2012. The numerical work presented in this paper was done on 
high computing facility at IISER Mohali.

\end{document}